\documentstyle[epsf]{l-aa}     
 
\begin{document}
 
   \thesaurus{08      
              (16.4;  
               03.2;  
               05.3;  
               09.2); 
             }
   \title{Atmospheric abundances in post-AGB candidates of intermediate temperature\thanks{Based on observations obtained at the Haute-Provence Observatory, France.}}
\markboth{Abundances}{Arellano Ferro} 
   \author{A. Arellano Ferro$^1$, Sunetra Giridhar$^2$, P. Mathias$^3$.}
   \institute{$^1$Instituto de Astronom\'{\i}a, Universidad Nacional Aut\'onoma de M\'exico,\\
               Apdo. Postal 70-264,   
M\'exico D.F. CP 04510; \\armando@astroscu.unam.mx \\
              $^2$Indian Institute of Astrophysics,  Bangalore 560034, India\\giridhar@iiap.ernet.in\\
$^3$ Observatoire de la C\^ote d'Azur, D\'epartment Fresnel, UMR 6528, B.P. 4229, F-06304 Nice Cedex 04, France\\mathias@obs-nice.fr}
 
\date{Received xxxxx; accepted xxxxx}
\maketitle
 
   \begin{abstract}
%
Detailed  atmospheric abundances have been calculated for a 
sample of A-G supergiant stars with IR fluxes and/or 
high galactic latitudes.  HD 172481 and HD 158616 show clear indications
  of being  post-AGB stars that have experienced third dredge-up.
HD 158616  is carbon-rich while the abundance pattern of
HD 172481 and its large Li enhancement gives support to the hot 
bottom burning scenario that
explains  paucity of carbon-rich stars among AGB stars. 
 HD 172324 is very likely a hot post-AGB star that shows a strong 
carbon deficiency. 
 HD 725, HD 218753 and HD 331319 also appear
 to be evolved objects between the red giant and the AGB. 
HD 9167, HD 173638 with a few  
 exceptions, reflect solar abundances and no signs of post red giant evolution.
They are most likely young massive disk supergiants. Further analysis of proto-Planetary Nebula HDE 341617
 reveals that He lines show signs of velocity stratification.   
 The emission lines have weakened 
 considerably since 1993. 
  The envelope expands at 19 km~s$^{-1}$ relative to the star. Atmospheric abundances, 
evolutionary tracks and isochrones are  used to estimate masses and 
ages of all stars in the sample. 
      \keywords{Stars: post-AGB; Stars: chemically peculiar; Stars: evolution;
Stars:individual HD 725, HD 9167, HD 158616,
HD 172324, HD 172481, HD 173638, HD 218753, HD 331319, HDE 341617.}
   \end{abstract}
 
%
 
\section{Introduction}
The  atmospheric chemical composition of post-asymptotic giant branch (post-AGB)
 stars  and circumstellar
environments is determined by nucleosynthesis and  
dredge-up events at the late AGB phases.
At AGB, the star has a C$-$O core surrounded by helium and hydrogen
 burning shells above which lies a deep convective envelope.
 The thermally pulsating phase (TP-AGB), 
 though much shorter than early AGB phase
 (E-AGB), is responsible, through considerable mass-loss, for the ejection of large fraction of carbon
 and $s$-process elements into the ISM. At this phase, thermal pulses
 are caused by instabilities in the He-burning shell first discovered 
 by Schwarzschild \& H\"arm (1965) and Weigert (1966).
 The excess energy generated by He shell flashes is transported 
by convection over the region that extends from the base of He-burning
shell and the hydrogen-helium discontinuity. The He shell instabilities 
strongly influence the chemical composition of the convective envelope. The 
 expansion and cooling of the intershell layers during  a powerdown phase
 of the He shell flash causes the deepening of the convective envelope into regions 
 containing the products of partial He-burning. The $^{12}$C, $^{19}$F and the
  $s$-process elements are mixed into the outer envelope causing 
 abundance variations at the surface of these stars.
 This process, known as the third
dredge-up (TDU), is able to explain the formation of carbon stars
(Busso, Gallino \& Wasserburg 1999), Wallerstein et al. (1997), Mowlavi
 (1999). At AGB the star is constantly losing mass, but a final phase
 of enhanced mass-loss by the superwind is believed to terminate the AGB
 phase producing a planetary nebula.
Therefore, studying the chemical composition of the
atmospheres and envelopes of evolved stars with IR fluxes, 
one expects to identify
post-AGB stars and to provide important observational
constraints for the theoretical work on nucleosynthesis, internal structure
and mass-loss in evolved intermediate and low-mass stars.

Post-AGB stars, as they evolve across the H-R diagram
 towards the white dwarf stage,
form families of rather exotic objects like the R CrB stars,
 other subgroups of
 H-deficient and He-rich stars, planetary nebulae etc. 
 In the H-R diagram, they populate the region generally occupied by 
  massive young supergiants evolving redwards 
from the main sequence,
and  having similar temperatures and luminosities. To differentiate
the massive and young stars from the highly evolved low-mass post-AGB stars,
 detailed atmospheric abundance analysis is crucial.
    
Chemical analysis of high galactic latitude A-F supergiants have led to the
discovery of many  
 interesting post-AGB stars such as 
HR 7671 (Luck et al. 1990), HR 4912 (Lambert et al. 1983),
HR 4114 (Giridhar et al. 1997) or of selected IRAS sources
such as IRAS 22223+4327 and IRAS 04296+3429 (Decin et al. 1998).
 However, most of these high  galactic latitude  
 stars are   field stars of unknown distances.
 It is therefore likely that a significant fraction of them could 
 possibly turn out to be disk objects of nearly solar compositions.
 A search of post-AGB stars among high galactic latitude stars could be
 more rewarding if we put the
 additional constraint of IR detection.
 The wavelength dependence of IR fluxes and also the detection of submillimeter
 fluxes could give valuable information on the circumstellar
 matter surrounding the evolved star.
 The IRAS two colour diagrams such as those published by Olnon et al.
 (1984), van der Veen and Habing (1988) etc., are extremely useful in
 separating stars with different kinds of envelopes.

In this  study,  we have undertaken the abundance analysis of a
selected  sample of stars likely to be
post-AGB stars. From the published lists of  high galactic latitude 
 stars ($b > 20^o$) (e.g. Bidelman 1990  and others)
 we chose the ones with known infrared fluxes. Among them, the ones
falling into the regions VIa and VIb  of figure 5b of 
van der Veen \& Habing (1988) were preferred as they were more likely  
to be post-AGB stars. We have also included a few objects belonging
 to the regions  IIIa and IIIb  that are likely to be 
 evolved stars with oxygen-rich envelopes. 

 The hot star HD 172324 was also included in spite of not being an 
IRAS source since it has high radial velocity ($-$110 km~s$^{-1}$)
 and very complex structures in hydrogen line profiles. It 
 appeared to be a possible hot post-AGB star similar to those 
investigated by  Conlon et al. (1993a;b).

  A search for post-AGB stars among  supergiant-like stars of 
 high galactic latitude is expected to be  more efficient   
since the possibility of forming stars at truly large distances from the
galactic plane is low.

This program is also aimed at providing calibrators for photometric 
empirical calibrations of atmospheric abundances
(Arellano Ferro \& Mantegazza 1996), temperatures, 
and gravities in  particular,   since
gravities are better  determined from the  ionization equilibrium.

This paper is organized in the following way: Sect. 2 describes the 
observations and data reduction;  Sect. 3 discusses the methodology 
 of abundance calculation; Sect. 4 gives an account of the 
sources of uncertainty
in the derived abundances; in Sect. 5 the results are given and discussed for
each star; 
in Sect. 6  these results are discussed in terms of the evolutionary status
of each star while in Sect. 7 we summarize our results.

\section{Observations and data reduction} 

The observational material for this work was obtained  during 
July 6 - July 12, 1999  with the 1.93m telescope
of the Haute-Provence Observatory (OHP), which is equipped with the high resolution (42,000) echelle 
spectrograph ELODIE. Details about the  performance and characteristics of the 
instrument have  been thoroughly described by Baranne et al. (1996).
 We have used one spectrum of HD 172324 taken on June 1995 with the Sandiford echelle
 spectrograph at 2.1m telescope of McDonald Observatory.
 This instrument giving a  resolution of 50,000 has been described in 
 McCarthy et al. (1993).
 One spectrum of HD 172481 was obtained with the 2.7m 
 2dcoud\'e echelle spectrograph
 described in Tull et al. (1995).
 These spectra were reduced using spectroscopic
 data reduction tasks available in the IRAF package.

\subsection{The sample}

Table 1 contains the list of stars studied in this work, their spectral 
types, magnitudes, galactic positions and, when available, the IRAS infrared fluxes. 


\begin{table*}
\caption{Basic data and IRAS fluxes of sample stars}                                                                 
\begin{center}
\begin{tabular}{llrrrrrrrr}
\noalign{\smallskip}                                                              
\hline       
\noalign{\smallskip}                                                              
\noalign{\smallskip}
\multicolumn{1}{c}{Star}&
\multicolumn{1}{c}{Sp.T.}& 
\multicolumn{1}{c}{$V$}& 
\multicolumn{1}{c}{$l$}& 
\multicolumn{1}{c}{$b$}&
\multicolumn{1}{c}{IRAS}& 
\multicolumn{1}{c}{12$\mu$}&  
\multicolumn{1}{c}{25$\mu$}&  
\multicolumn{1}{c}{60$\mu$}&  
\multicolumn{1}{c}{100$\mu$}\\
\noalign{\smallskip}
\multicolumn{1}{c}{}&
\multicolumn{1}{c}{}& 
\multicolumn{1}{c}{$(mag.)$}& 
\multicolumn{1}{c}{$(^o)$}& 
\multicolumn{1}{c}{$(^o)$}&
\multicolumn{1}{c}{}& 
\multicolumn{1}{c}{$(Jy)$}&  
\multicolumn{1}{c}{$(Jy)$}&  
\multicolumn{1}{c}{$(Jy)$}&  
\multicolumn{1}{c}{$(Jy)$}\\
\noalign{\smallskip}  
\hline  
\noalign{\smallskip}  

HD 725& F5Ib-II &7.08&117.56 & $-5.19$&00091+5659&.36& .25& .40& 14.43\\
HD 9167&F1II&8.19&127.73 & $-0.97$&01285+6115&.47& .25& .40& 8.14\\
HD 158616& F8 &9.69&13.23& $+12.17$&17279$-$1119&3.52& 2.90& 1.60& 1.98\\
HD 172324&B9Ib&8.16&66.18 &$+18.58$&&&&&\\
HD 172481&F2Ia0&9.09&6.72& $-10.37$&18384$-$2800& 5.41&5.22& .59&1.85\\
HD 173638& F2Ib-II&5.73&23.38 & $-3.56$&18439$-$1010&1.41& .39&.67&55.03\\
HD 218753&A5III&5.69&110.28 & $-1.02$&&&&&\\
HD 331319&F3Ib&9.50&67.16&$+2.73$&19475+3119&.54& 37.99& 55.83& 14.76\\
HDE 341617&A5&9.40&50.67&$+19.79$&18062+2410&3.98&19.62 &2.90 &1.00\\
            \noalign{\smallskip}  
            \hline  
            \noalign{\smallskip}  
\end{tabular}
\end{center}
\end{table*}  

\subsection{Spectroscopic reductions}

All spectra were bias-subtracted and flat-field corrected using 
standard OHP procedures described by Baranne et al. (1996).
 The spectra were wavelength calibrated with Th-Ar hollow cathode
 lamp spectra taken after each stellar exposure. More than one
 exposure was taken and spectra were combined to attain $S/N$ of at least 50.
 The ELODIE spectrograph gives a resolution of 42,000 and 
the wavelength coverage goes from 3906 to 6811~\AA ~    
in 67  echelle orders
 with some overlaps in adjacent orders.

 The equivalent widths were measured using the 
{\it splot} task of the IRAF package and their accuracy is  generally
better than 10\% for spectra with $S/N$ ratio larger than 50.
 We generally restricted ourselves to unblended weak features and 
avoided using lines stronger than 200~m\AA.

\section{Abundance Analysis}  

 We have used ATLAS9 (Kurucz 1993) model atmospheres as an input to the
 1997 version of LTE line synthesis program MOOG first described in 
 Sneden (1973). The procedure assumes plane-parallel atmospheres,
 hydrostatic equilibrium and LTE.  
 The oscillator strength or $gf$ value is an important atomic datum
 that affects the abundance calculations.
 For elements C, N and O we used $gf$ values from Wiese, F\"uhr \& Deters (1996).
For Fe I, the values were taken in order of preference from;  Table A1 of
 Lambert et al. (1996), Luck's compilation (1996; private communication), and 
Giridhar \& Arellano Ferro (1989).
For Fe II lines we used the  
Table A2 of Lambert et al. (1996), Giridhar \& Arellano Ferro (1995) and  Luck's  compilation (1996; private 
communication) .

For elements other than Fe, the large compilation by 
Luck (1996; private communication) was preferentially used
and for some heavy $s$-process elements $gf$ 
 values were taken from the work of Th\'evenin (1989; 1990).

\subsection{ Determination of atmospheric parameters}

 In addition to abundances, the line strengths are strongly affected by
 atmospheric parameters like the  effective temperature ($T_{\rm eff}$), gravity (log $g$) 
and turbulent velocity ($\xi_{t}$). 
 It is  therefore necessary to determine these parameters before 
 using line strengths for abundance determinations. 

\subsubsection{Effective Temperature}

 Temperature calibrations  exist that are valid
  for specific spectral type ranges. These  methods are
not only useful but also complement the spectroscopic efforts by providing initial 
 values for the atmospheric  parameters  for   calculating
the atmospheric abundances. Here we briefly describe some of those
calibrations that have been used for the stars of our sample. 
Later in Sect. 5, comparison with the finally adopted spectroscopic results is made in the discussion of each individual star.

   Firstly, a rough estimate of $T_{\rm eff}$ can be made from
the given spectral type and the calibration of Schmidt-Kaler (1982).
However, more accurate values can be obtained from precise 
photometric colours. For our sample, we have used $uvby \beta$ 
photometric data and our
own unpublished calibrations for F-G stars (HD 725, HD 9167, HD 172481).
 In addition, the calibration of
Napiwotzki, Sh\"onberner \& Wenske (1993) for  hotter stars like  HD 172324,
 the 13-colour photometric system and the calibration of 
Bravo Alfaro et al. (1997), and the                
   Geneva colour indices and the calibration of Cramer \& Maeder (1979)
  were used as appropriate.    These empirical calibrations provide 
$T_{\rm eff}$ 
with accuracies of $\pm 500$ K or better and serve as excellent starting
values that are further refined by 
  spectroscopic approaches. Any  drastic difference
 between the two approaches  deserves attention.

Yet another independent approach to estimate $T_{\rm eff}$  is from the
Balmer lines profiles fitting (e.g. Arellano Ferro 1985; Venn 1995a). Theoretical Balmer profiles for  grids  of model atmospheres 
 have been calculated by Kurucz (1993). For stars of  
intermediate temperature, this method does not 
 lead to unique $T_{\rm eff}$,
 but rather it defines loci of possible temperatures and gravities.
On the other hand, similar loci can be found 
from species  where two states of ionization are well represented.
Again, the solution is not unique  but rather  a locus on the 
$T_{\rm eff}$ - $\log g$ plane is defined for each element.  
This approach will be illustrated in Fig. 3 for H$_\gamma$, H$_\delta$, Mg and 
Si for the star HD 218753. The above solution is of special
 importance  for hot stars where no Fe I lines are present.
For hotter stars, lines of Fe I are not only very weak  but are also 
influenced by non-LTE effects.  For A-F supergiants non-LTE effects could cause errors in the range of 0.2 to 0.3 dex in the iron abundance
derived using Fe I lines (Boyarchuck et al. 1985). For example, in 
 the well-known star Vega (A0V), the neglect of departure 
from LTE for Fe I lines leads to the underestimation of Fe abundance by 0.3 dex (Gigas 1986).

For stars cooler than 7500 K the lines of Fe I and Fe II with wide range 
in line strengths and lower excitation potential are adequate for estimating the
effective temperature, microturbulence and gravity for any given star.

The finally adopted $T_{\rm eff}$ for our stars is that for which abundance
consistency is obtained from neutral and ionized lines of well represented species such as Fe, Ti and Cr.

\begin{table*} 
\caption{Physical and dynamical parameters derived for program stars}                                                                 
\begin{center}
\begin{tabular}{lccccccc}
\noalign{\smallskip}                                                              
\hline       
\noalign{\smallskip}                                                              
\noalign{\smallskip}
\multicolumn{1}{c}{Star}&    
\multicolumn{1}{c}{$T_{\rm eff}$}&  
\multicolumn{1}{c}{$\log g$}& 
\multicolumn{1}{c}{$\xi_{t}$}&
\multicolumn{1}{c}{$V_r (hel)$ }& 
 \multicolumn{1}{c}{$V (LSR)$}&
 \multicolumn{1}{c}{$\sigma_{V_r}$}&
\multicolumn{1}{c}{log $(L/L_{\odot})$}\\
\noalign{\smallskip}
\multicolumn{1}{c}{}&    
\multicolumn{1}{c}{$(K) $}&  
\multicolumn{1}{c}{}& 
\multicolumn{1}{c}{$(km~s^{-1})$}&
\multicolumn{1}{c}{$(km~s^{-1})$ }& 
 \multicolumn{1}{c}{$(km~s^{-1})$}&
 \multicolumn{1}{c}{$(km~s^{-1})$}&
\multicolumn{1}{c}{}\\
            \noalign{\smallskip}  
            \hline  
            \noalign{\smallskip}  

HD 725& 7000 & 1.0 & 4.65 &$-56.9$&$-48.8$& 0.8&4.2\\
HD 9167&7250 & 0.5 & 4.20 &$-45.7$ &$-40.2$&1.1 &4.2\\
HD 158616& 7300 &1.5&4.6& $+68.8$ &$+78.6$&2.1&4.5\\
 & &&& $+63.7$ &$+78.5$&1.5&   \\
HD 172324&11000& 2.5&5.0&$-126.1$ &$-106.4$&1.1&2.7\\
 &11500&2.5&7.5&$-117.3$&$-97.5$&2.5&   \\
HD 172481&7250& 1.5&4.60&$-73.1$&$-76.9$&1.8 &4.1,4.5\\
&7250& 1.5&5.10&$-84.4$&$-74.0$&2.1&   \\
HD 173638& 7500&1.5&4.30&$+11.6$&$+26.9$&1.2 &4.5\\
HD 218753&8000&2.0& 3.35& $+3.2$& $+13.9$&1.0&1.6\\
HD 331319&7750&1.0 & 5.35&$ -2.5$&$+15.8$&1.5 &4.5\\
HDE 341617&23000&3.0&15.00&+67.9&+87.8&2.9 &4.6\\
            \noalign{\smallskip}  
            \hline  
            \noalign{\smallskip}  
\end{tabular}
\end{center}
\end{table*}

\subsubsection {Gravity}

  A good discussion of atmospheric  parameters determination for A type
 stars can be found in Venn (1995a), who points out 
 that H$_\gamma$ and 
 H$_\delta$, being very sensitive to temperature and gravity
 in A type stars, provide a locus of possible temperature-gravity pairs,
and that ionisation equilibrium of Mg I and 
 Mg II gives another useful locus of temperature-gravity pair as
 the non-LTE effects are expected to be very small in magnesium lines.
 Since ionisation equilibrium of Si I and Si II give a temperature-gravity
 pair very similar to that given by Mg I and Mg II, the latter
 can also serve as yet another indicator of these parameters. 
 Intersection  of the above mentioned loci could lead to reliable
 temperature  and gravity for each star, as demonstrated in Fig. 3.

 The hydrogen lines were distorted in many of the  program stars 
 due to underlying emission,
 and therefore could not be used  to derive temperature-gravity loci. 
We used excitation equilibrium of Fe I lines to get a preliminary estimate of 
$T_{\rm eff}$. It was followed by  ionisation equilibrium               
 of Mg~I/Mg~II, Si I$/$Si II and Cr~I$/$Cr~II 
 to arrive at a satisfactory estimate of $T_{\rm eff}$ and $\log g$. 
 For HD 725,
 HD 9167, HD 158616, HD 172481 and HD 173638 the excitation equilibrium of
 Fe I lines (requiring derived abundances to be independent
 of the lower excitation energy of the lines) gave very good estimates 
 of the temperature which were further   verified using  lines 
 of other species, as mentioned above. Similarly, for gravities,
 the values giving a good consistency for neutral and ionised  Mg,
 Ti, Cr and Fe were adopted. 

The star HD 172324 required an altogether different approach as  
described in Sect. 5.4.

\subsubsection {Microturbulence velocity}

 All our program stars turned out to be hotter than 
 7000 K (see Table 2). For hotter stars, Fe I lines were difficult to
 measure. The Fe II lines on the other hand, had good range in 
 equivalent widths. We therefore relied upon 
  Fe II lines to derive microturbulence. The microturbulence was 
 derived by requiring that weak, medium and strong lines give a consistent
 value of abundance.

 The final atmospheric parameters derived for the program stars are
given in Table 2,
along with their radial velocities relative to the 
Sun and to the Local Standard of Rest (LSR). 
The log $(L/L_{\odot})$ values in Table 2 were estimated from the 
effective temperatures determined spectroscopically,
 and the calibration of
Schmidt-Kaler (1982). For HD 172481 a red spectrum obtained at 
McDonald Observatory in May, 2000,
allowed us to measure the three components of the OI feature near 7774~\AA . 
The  combined equivalent width W(7774) = 1.3 $\AA$ and the 
calibration of Arellano Ferro, Giridhar \& Goswami
(1991) lead to $M_v$ = $-$5.6 or log $(L/L_{\odot})$ = 4.1, 
which is in good agreement with the value 4.5 
obtained from Schmidt-Kaler's calibration. For homogeneity we have adopted
 the latter value for our discussion  about the evolutionary status
 of this object in Sect. 6.

\section{Uncertainties in the Elemental Abundances }

The uncertainties in the derived abundances are caused by errors in the determination
of the atmospheric parameters, in the equivalent width measurements, and
 also in the quality of oscillator strengths.
   For spectra with  $S/N$ ratios larger than 50
 the errors in the equivalent widths are between 5 and 8\%.
 The errors in $gf$ values vary from element to element. For Fe I lines, 
 experimental values of good accuracies (better than 10\%) do exist, 
 for other Fe-peak
 elements the range in errors could be within  10 to 25\%. For heavier elements,
 particularly for $s$-process elements, the errors could be larger than 25\%.
 For the stars HD 725, HD 158616, HD 172481, HD 173638, HD 218753 
 and  HD 331319 we could measure a very large number of unblended lines, and the estimated errors in $T_{\rm eff}$, $\log g$ and  $\xi_{t}$
 are $\pm$ 200 K, $\pm$0.25 and $\pm$0.2 km~s$^{-1}$, respectively.
 The sensitivity of the 
  derived abundances to changes in the model atmospheric parameters
 are described in Table 4 of Gonzalez et al. (1997) for two RV Tau stars.
 We have used the same grid of  atmospheric models and the same database 
 for line oscillator strengths, hence the procedure will not be repeated here.
 For  the  Fe-peak elements we could measure a sufficiently large number 
  of lines and the $gf$ values used being of good quality, we expect these
 abundances to 
   be accurate within 0.2 to 0.25 dex. For heavier elements, particularly
 the $s$-process elements, 
the uncertainty  could  be above 0.3 dex. Similarly for light elements like
 oxygen where few lines are available the uncertainty could be above 0.3 dex.
 For  HD 172324 and HDE 341617, we will discuss the uncertainties 
 in  their respective sections. 
 
\section {Results}
In what follows, we present derived elemental abundances for individual stars.
\subsection{HD 725}

This star is an IRAS source (00091+5659). It was classified as F5Ib-II 
by Griffin \& Redman (1960) suggesting a
 temperature of $\sim$ 6900 K if we follow the calibration of 
Schmidt-Kaler (1982).

Using the $uvby$ photometry of Perry (1969), Olsen (1983), Hauck \& Mermilliod
(1998), reddening free colours and our own 
unpublished calibrations for F-G supergiant stars, we estimated 
$T_{\rm eff}$ = 6900 K. Balmer line fitting was  not performed because the profiles 
display  complex structure and
underlying emission is suspected.

\begin{table} 
\caption{Elemental Abundances for HD 725}                                                                 
\begin{center}
\begin{tabular}{lcccrc}
\noalign{\smallskip}                                                              
\hline       
\noalign{\smallskip}                                                              
\noalign{\smallskip}
\multicolumn{1}{l}{Species}&  
\multicolumn{1}{c}{$\log \epsilon_{\odot}$}&  
\multicolumn{1}{l}{[X/H]}&  
\multicolumn{1}{l}{s.d.}&  
\multicolumn{1}{c}{N}&  
\multicolumn{1}{r}{[X/Fe]}\\
           \noalign{\smallskip}  
            \hline  
            \noalign{\smallskip}  
C I& 8.55 &$-0.38$ &$\pm$0.03 & 3 &$-0.08$\\
Na I& 6.32 & $+0.21$&$\pm$0.08 &3 &$+0.51$  \\
Mg I& 7.58 &$-0.18$ &$\pm$0.25 & 4  & $+0.12$ \\
Si I& 7.55 &$+0.05$ &$\pm$0.03 & 2  &$+0.35$\\
Si II& 7.55 &$+0.12$ && 1  &$+0.42$\\
S I& 7.21 & $-0.10$&$\pm$0.17 &3 & $+0.20$ \\
Ca I& 6.35 &$-0.16$& $\pm$0.10 &12 & $+0.14$\\
Sc II& 3.13 & $-0.03$& $\pm$ 0.23  &7 & $+0.27$\\
Ti II&4.98 & $-0.31$&$\pm$0.08 & 6& $-0.02$\\
Cr I&5.67 & $-0.16$&$\pm$0.17 & 7 & $+0.14$\\
Cr II&5.67 & $-0.19$&$\pm$0.16 & 11 & $+0.11$\\
Mn  I&5.39 & $-0.27$&$\pm$0.12 & 6 & $+0.03$\\
Fe  I&7.51 & $-0.26$&$\pm$0.14 &  51 & \\
Fe II&7.51 & $-0.33$&$\pm$0.15 & 11  & \\
Ni I& 6.25 & $+0.04$ & $\pm$0.14&3 &$+0.33$ \\
Y II& 2.23 & $+0.12$ &$\pm$0.15& 3 &$+0.42$ \\
Zr II&2.60 & $-0.07$ &$\pm$0.13 & 2 & $+0.22$ \\
Ba II& 2.13 & $+0.15$&&1& $+0.45$\\
Ce II& 1.58 & $-0.17$&$\pm$0.13 &2& $+0.13$\\
            \noalign{\smallskip}  
            \hline  
            \noalign{\smallskip}  
\end{tabular}

Notes -- The solar abundances are taken from  Grevesse, Noels
 \& Sauval (1996).\

-- N is the number of lines included in the calculation.\\
\end{center}
\end{table}   

The finally adopted parameters are $T_{\rm eff}$ = 7000 K, $\log g$ = 1.0 and 
$\xi $$_{t}$ = 4.65 km~s$^{-1}$. We relied upon Fe II lines for calculating
 microturbulence velocity. A large number of Fe I lines were measured
and used in estimating $T_{\rm eff}$. Also 
  neutral and ionized lines
of Ti and Cr were employed to derive a satisfactory pair of $T_{\rm eff}$ 
and $\log g$ values.
A competing solution  could have been  
$T_{\rm eff}$ = 7200 K, $\log g$ = 1.5 but the adopted parameters gave 
 marginally better consistency in the abundances of neutral and ionized lines.
 HD 725 appears to be very marginally metal-poor (Table 3) and
 it has a moderately high radial velocity of $-$57 km~s$^{-1}$.
The elemental abundances relative to Fe, i.e. [X/Fe] = [X/H] - [Fe/H]
\footnote{using the spectroscopic notation
[X/H]= log [(X/H) $-$ (X/H)$_{\odot}$]}, 
given  in the last column of
Table 3 and subsequent tables, were calculated adopting the average value
of [Fe/H] from Fe I and Fe II lines.
For HD 725 the [X/Fe] values  are similar
 to those in normal unevolved stars for many elements. [Na/Fe] = +0.5  dex  
 indicates relative enrichment of Na which is a well-known feature of
 A-F supergiants (Takeda \& Takada-Hidai 1994, Venn 1995a, b).
 But Na I abundances are likely to be affected by non-LTE effect.
Non-LTE analysis of Na I lines has been done by Gigas (1986),
  Takeda \& Takada-Hidai (1994) and others. The errors introduced 
 by the neglect of non-LTE becomes more severe for higher temperatures.
 According to Takeda \& Takada-Hidai (1994), the $\triangle$log
 $\epsilon$ in Na I lines at 5682, 5688, 6154 and 6160~\AA~ is $-$0.09, $-$0.10,
 $-$0.07 and $-$0.07 dex respectively at temperature 7500 K. At temperature
 7000 K the correction is 0.01 dex smaller for all lines than the
 values mentioned above. According to Gigas (1986) the non-LTE
 correction could be +0.1 to +0.2 dex. The use of [Na/Ca]
 instead of [Na/Fe] is recommended by Lambert (1992) for LTE
 calculations. The [Na/Ca] of +0.4  dex  found from our analysis shows 
 that Na enrichment appears to be real. The Na might have been synthesized
 in the H-burning region where the NeNa cycle might operate together 
 with the CNO cycle. The suggestion that the proton capture on $^{22}$Ne
 could lead to enhancement of $^{23}$Na is followed up  by Langer, 
 Hoffman \& Sneden (1993).  These authors used a nuclear reaction network 
 to examine the changes in abundances caused by proton capture at $T_{9}$
 =0.040.  Mixing of this region just below the oxygen shell 
 over a timescale of 30,000 yr would cause enhancement of 
  $^{14}$N, $^{23}$Na and $^{27}$Al
   at the expense of  $^{16}$O, $^{22}$Ne, $^{25}$Mg and
 $^{26}$Mg. These authors also suggested that the abundant $^{20}$Ne could
 also be transformed into $^{23}$Na on longer timescales, and
 pointed out that the depletion of $^{25}$Mg and $^{26}$Mg would not modify
the Mg abundance.
 Globular cluster giants are known to display Na $-$ O anticorrelation
 as reported by Sneden et al. (1991) and Kraft et al. (1992).
  Since our spectrum
 does not go to  wavelengths longer than 6800~\AA ,
 we could not measure N I abundance nor could we measure the O abundance. 
 In metal-poor stars, relative enrichment of $\alpha$ elements is
 to be expected but the effect becomes evident for [Fe/H] $<$ $-$0.5 dex.
 At [Fe/H] $-$0.2 to $-$0.3 dex the spread in observed abundances is
 large and a large number of stars have [$\alpha$/Fe]$\sim$0.
 For HD 725, the $\alpha$ element Si shows small enhancement
 but the number of lines used are woefully small to make a 
 definite claim.

 Among Fe-peak elements, Ni, represented by 3 Ni I lines,
 appears to be enriched, with [Ni/Fe]
 of +0.3 dex. HD 218753 and HD 173638 of our sample show a
 positive [Ni/Fe] but the value is not above abundance errors.
 As such, for HR 725 the value is slightly above twice the 
standard error, nevertheless 
 a more extensive analysis based on a larger number of lines  is
 required to see if the enrichment is real. 
 Luck \& Bond (1983, 1985) have reported [Ni/Fe]$\sim 0 $ for 
 metal-poor star. Wheeler, Sneden \& Truran (1989)
 on the other hand report very large scatter in [Fe/H] vs. [Ni/Fe] 
 relation. These authors suggest that non-zero [Ni/Fe] can be 
 observed at all metallicities.  

 Another interesting finding is [Y/Fe] of +0.4  dex. 
 The three Y II lines used are quite well separated and have good
 estimates of oscillator strengths, 
 but out of the three lines, one  is a little
 strong. The same is true for the Ba II line used. 
 Mild enrichment of  $s$-process elements in the moderately metal-poor 
 star HD 70379 has already been reported (Reddy 1996). Hence, our average
 [s/Fe] ratio  +0.3  does not come as a surprise. In addition, 
     its radial velocity
 of $-$57 km~s$^{-1}$ lends support to our view that it is a
 low-mass  evolved object.
 These results are  highly suggestive though not conclusive indicators of
 evolution beyond the red giant branch.
However, making use of its proper motions and 
parallax from the Hipparchos catalogue we have calculated 
the galactocentric velocities  $\Pi = -32.6$ km~s$^{-1}$,
 $\Theta = +190.3$ km~s$^{-1}$ and
$Z = +5.4$ km~s$^{-1}$ that indicate a space velocity of 
193.2 km~s$^{-1}$
with a small pitch angle of 9.7$^{o}$ relative to the 
circular orbit and on the
galactic plane. This indicates that the star is at a lower galactic
latitude than the Sun and placed in
a mildly eccentric orbit.                

\subsection{HD 9167}

This star has infrared flux and is an IRAS source (01285+6115). It appears to be another moderately metal-poor star (Table 4).
 We did not find any remarkable abundance peculiarity for this object.
 With radial velocity of $-$45 km~s$^{-1}$ and [Fe/H] of $-$0.3 dex,
 one could be optimistic of seeing positive [$\alpha$/Fe].
 Unfortunately, we could not measure good Si I and Si II lines and Mg, Ca and Ti
 do not show any enrichment.  But on the  other hand, for stars with 
 [Fe/H] in the
 range of 0.0 to $-$0.5 dex, the observed abundance ratios of $\alpha$-process
elements have very 
 large scatter, hence finding evolutionary changes is 
almost as probable as not finding them. Likewise HD 725, rather large 
radial velocity implies a space velocity of 208.2 km~s$^{-1}$
with a small pitch angle of 5.7$^{o}$ relative to the circular orbit and on the
galactic plane. The orbit is even less eccentric than that of HD 725 and it is 
at a lower galactic latitude than the Sun.

\begin{table} 
\caption{Elemental Abundances for HD 9167}                                                                 
\begin{center}
\begin{tabular}{lcccrc}
\noalign{\smallskip}                                                              
\hline       
\noalign{\smallskip}                                                              
\noalign{\smallskip}
\multicolumn{1}{l}{Species}&  
\multicolumn{1}{c}{$\log \epsilon_{\odot}$}&  
\multicolumn{1}{l}{[X/H]}&  
\multicolumn{1}{l}{s.d.}&  
\multicolumn{1}{c}{N}&  
\multicolumn{1}{r}{[X/Fe]}\\
           \noalign{\smallskip}  
            \hline  
            \noalign{\smallskip}  
Mg I& 7.58 &$-0.37$ &$\pm$0.15 & 2  & $-0.04$ \\
Ca I& 6.35 &$-0.22$& $\pm$0.09 &6 & $+0.12$\\
Sc II& 3.13 & $-0.28$& $\pm$ 0.09  &3 & $+0.05$\\
Ti II&4.98 & $-0.48$&$\pm$0.17 &  10& $-0.15$\\
Cr I&5.67 & $-0.19$& & 1 & $+0.15$\\
Cr II&5.67 & $-0.27$&$\pm$0.21 & 9 & $+0.07$\\
Mn  I&5.39 & $-0.35$& & 1 & $-0.02$\\
Fe  I&7.51 & $-0.32$&$\pm$0.16 &  29 & \\
Fe II&7.51 & $-0.35$&$\pm$0.21 & 10  & \\
Y II& 2.23 & $-0.23$ &$\pm$0.27& 3 &$+0.11$ \\
Ba II& 2.13 & $-0.22$ & $\pm$0.09 & 2 &$+0.12$ \\
            \noalign{\smallskip}  
            \hline  
            \noalign{\smallskip}  
\end{tabular}

Notes -- same as Table 3.\\
\end{center}
\end{table}   

\subsection{HD 158616}

 This star  is an IRAS source (17279$-$1119) with significant infrared fluxes 
at shorter wavelengths. 

The spectral type as listed in the SAO catalogue is F8, which suggests a 
temperature between 6100-6200 K (Schmidt-Kaler 1982), 
depending upon the luminosity class. No photometric data 
are available and hence no other temperature estimate was made.
Two spectra were obtained for this star at the OHP. The $S/N$ for these two 
spectra are 34 and 53. Since the spectrum from  July 7 is better, 
it was decided to use it to determine $T_{\rm eff}$, 
$\log g$ and $\xi$$_{t}$, complete the abundance analysis and 
then simply use the same parameters on the lower $S/N$ spectrum
from July 6 to  verify the abundance pattern. The results are given in Table 5. The
second entries for each species  are for the lower $S/N$ spectrum. One can see
that the results from both spectra are in a good agreement.

\begin{table} 
\caption{Elemental Abundances for HD 158616}                                                                 
\begin{center}
\begin{tabular}{lcccrc}
\noalign{\smallskip}                                                            
\hline       
\noalign{\smallskip}                                                              
\noalign{\smallskip}
\multicolumn{1}{l}{Species}&  
\multicolumn{1}{c}{$\log \epsilon_{\odot}$}&  
\multicolumn{1}{l}{[X/H]}&  
\multicolumn{1}{l}{s.d.}&  
\multicolumn{1}{c}{N}&  
\multicolumn{1}{r}{[X/Fe]}\\              
            \noalign{\smallskip}  
            \hline  
            \noalign{\smallskip}  
C I& 8.55 &$-0.25$ &$\pm$0.20 & 4 &$+0.33$\\
&& $-0.23$&$\pm$0.14 &  4 &+0.35 \\
O I& 8.87 & $-0.54$&$\pm$0.02   &2 & $+0.04$ \\
&& $-0.55$& &  1 &+0.03 \\
Na I& 6.32 & $+0.05$&$\pm$0.01 &3 &$+0.63$  \\
&& $-0.05$& &  1 &+0.49 \\
Mg I& 7.58 &$-0.40$ &$\pm$0.11 & 3  & $+0.18$ \\
&& $-0.56$& &  1 &-0.02 \\
Si I& 7.55 &$+0.03$ &$\pm$0.17 & 3  &$+0.61$\\
Si II& 7.55 &$-0.06$ &$\pm$0.17 & 2  &$+0.52$\\
&& $+0.04$&$\pm$0.47  &  2 &+0.58 \\
S I& 7.21 & $+0.08$&$\pm$0.12 &6 & $+0.66$ \\
&& $+0.10$&$\pm$0.01  &  2 &+0.64 \\
Ca I& 6.35 &$-0.36$& $\pm$0.16 &12 & $+0.22$\\
&& $-0.34$&$\pm$0.31  &  9 &+0.20 \\
Sc II& 3.13 & $+0.00$& $\pm$ 0.27  &8 & $+0.58$\\
&& $-0.09$&$\pm$0.24  &  4 &+0.45 \\
Ti I&4.98 & $+0.08$&$\pm$ 0.20 & 2& $+0.66$\\
&& $+0.18$&&  1 &+0.72 \\
Ti II&4.98 & $-0.01$&$\pm$0.19 &  13& $+0.57$\\
&& $-0.13$&$\pm$0.11 & 8 &+0.41 \\
Cr I&5.67 & $-0.55$&& 1 & $+0.03$\\
Cr II&5.67 & $-0.39$&$\pm$0.14 & 14 & $+0.19$\\
&& $-0.47$&$\pm$0.22 & 7 &+0.07 \\
Mn  I&5.39 & $-0.17$&& 1 & $+0.41$\\
Fe  I&7.51 & $-0.58$&$\pm$0.15 &  39 & \\
&& $-0.61$&$\pm$0.18 &  20 & \\
Fe II&7.51 & $-0.57$&$\pm$0.19 & 19  & \\
&& $-0.51$&$\pm$0.19 &  7 & \\
Ni I& 6.25 & $-0.41$ & $\pm$0.08&4 &$+0.17$ \\
&& $-0.39$&          & 1 &+0.19 \\
Zn I& 4.60 & $-0.48$ & $\pm$0.02& 2 &$+0.10$ \\
&& $-0.40$&& 1 &+0.18 \\
Y II& 2.23 & $+0.37$ &$\pm$0.25& 4 &$+0.95$ \\
&& $+0.48$&$\pm$0.03 & 2 &+1.02 \\
Ba II& 2.13 & $+0.04$ & $\pm$0.22 & 3 &$+0.62$ \\
&& $+0.10$&$\pm$0.20 & 2 &+0.64 \\
La II& 1.21 & $-0.11$& &1&$+0.47$ \\
Ce II& 1.58 & $+0.13$&$\pm$0.22 &5&$+0.71$ \\
&& $+0.21$&& 1 &+0.75 \\
            \noalign{\smallskip}  
            \hline  
            \noalign{\smallskip}  
\end{tabular}

Notes -- same as Table 3.\

-- second entries are results from a lower $S/N$ spectrum\\
\end{center}
\end{table} 

 HD 158616  was also studied by Van Winckel (1995; 1997). While he
 had better data for C,N,O due to extended coverage in the long wavelength 
 region,  
for other elements, our spectra contained  more lines per element,
and more  elements are included.
 This star is  Fe-poor by a factor of 4 or so and
shows very clear indications of CNO processing and 
 enrichment of $s$-process elements.
 We get C/O$\sim$1 whereas Van Winckel (1995) got this
 ratio significantly  larger than one (see Table 14). 
His carbon abundance is based on  more lines in the 7100~\AA ~ 
 region. His  choice of $T_{\rm eff}$ is also hotter than our adopted value 
 that can also account for the higher carbon abundance derived. Van  
 Winckel (1995) found [N/Fe] of +0.2 dex. This  clearly shows that the star has 
 gone through CNO cycle and its products have been brought to the surface.
  The carbon enrichment could be caused by the helium-shell burning
 when oxygen is also manufactured.
 The  enhancement of Si and S could be 
 present in ISM from which the star is formed.  
  With the  [Fe/H] of  $-$0.6 dex, the star is moderately metal-poor
 and therefore it is possible that the  parent
 ISM might have received ejecta from type II SNe.
 Although the effect of departure from LTE has been discussed for 
 the light elements C,N,O and also for Mg by different investigators, we
 did not come across similar discussion on Si and S. Hence, at this stage,
 we chose not to offer any detailed explanation.
The most interesting feature of the derived abundances is definitely the
enrichment of $s$-process elements Y, Ba, La and Ce. 
 This star is undoubtedly a post-AGB star that has brought the products
 of helium burning as well as elements formed by $s$-processing to the surface.
 Our analysis covers the light $s$-process element Y (one of the
 three elements  referred as ls) and the
heavy $s$-process elements Ba, La and Ce,  representing  
 heavy (hs) $s$-process elements. We have made an estimate of
  [hs/ls]. We have not corrected  the ls-index  for the lack of
 light elements Sr and Zr since for light elements the odd-even effect is not
 strong  as pointed out by  Van Winckel \& Reyniers 2000).
We estimated [hs/ls] = $-$0.3  and [ls/Fe] of $+$1.0. 
 When plotted on the [hs/ls] vs. [ls/Fe] plot  of Busso et al. (1999), 
(their Fig. 7 giving theoretical predictions),
   we found $\tau_{0} \sim 0.22$ mbarn$^{-1}$.
  Incidentally, the data  point falls near the thick line
 which for solar metallicity would indicate C/O = 1 but for [Fe/H]
 of $-$0.6, it indicates a C/O $\sim$ 3, whereas we get C/O $\sim$ 1.
 This reduction of carbon abundance, while hs and ls indices point
 to larger C/O  added to small but significant enhancement of nitrogen, as
 reported by Van Winckel (1995), 
 strongly favour the hot bottom burning scenario described and discussed 
 in section 5.5.
More extensive coverage of $s$-process elements
 will enable a meaningful comparison with  third dredge-up  models
 developed by Straniero et al. (1995) and Busso et al. (1995).

 Another star of similar temperature and  showing a similar trend in 
abundances is HR 6144 (Luck et al. 1990),
however, in this star, the $s$-process enhancement is not so 
significant and C/O is less than 1.

\subsection{HD 172324}

This star has been classified as B9Ib by Morgan \& Roman (1950) 
 which suggests a temperature of 10280~K (Schmidt-Kaler 1982). The Str\"omgren colours (Hauck \& Mermilliod 1998) however
point towards higher temperature: 13100~K using the calibration of Napiwotzki, 
Sh\"onberner \& Wenske (1993). An independent temperature calibration of the 
Geneva photometric
system (Cramer \& Maeder 1979) gives a temperature of 13215~K.

The He I  lines have been used to estimate $T_{\rm eff}$ and luminosity class
of the star.
 Didelon (1982) gives very useful  plots of the dependence of 
many He I, Si II, Mg II, C II, O II and N II line strengths  on the           
 spectral type and luminosity class.
The equivalent widths in our spectrum for the  He I lines at 4120, 4143, 4387 
and 4471~\AA ~  very clearly suggest a 
spectral type of B9.  A luminosity class Ib was suggested by the strengths
 of Si II line at 4128~\AA ~, Mg II line at 4481~\AA ~ ~and C II
feature at 4267~\AA, this
is in good agreement with the spectral type above.

An estimate of $T_{\rm eff}$ has been made by requiring the He I 
 lines to give solar
abundance.     
 The best estimate is 11500~K. Given the class Ib, 
 $\log g$ must not be very different from 2.0. In any case,  at the high      
 temperature end the Kurucz (1993) models do not reach very low gravities.
 Fortunately, Mg I and Mg II lines are present in the OHP
 spectrum and Si II and Si III lines can be measured on the McDonald spectrum.
  Our derived $T_{\rm eff}$ and $\log g$ appear  to be  good estimates for
 the epoch of OHP spectrum. However there is indication that at the epoch at which  
 the McDonald spectrum was taken, the gravity was somewhat lower.
 
\begin{figure}  
\centering  
\mbox{\epsfxsize=2.9in\epsfbox{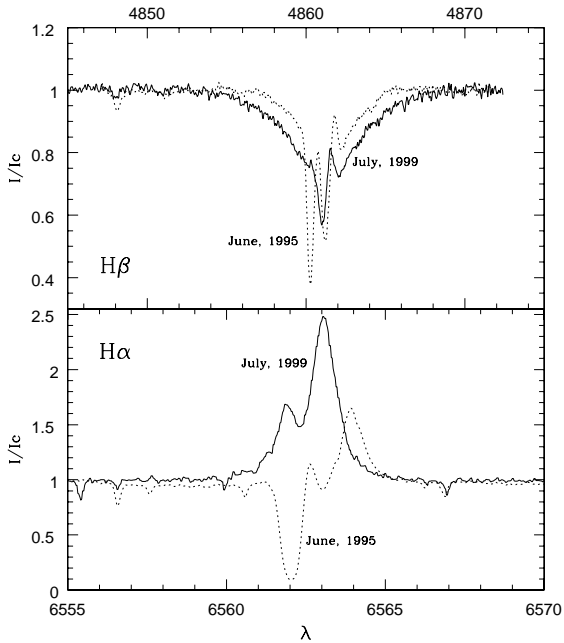}}  
\caption{H$_\alpha$ and H$_\beta$ variations exhibited by HD172324 in two spectra taken 4 years apart: June 1995 at McDonald and 
July 1999 at OHP.}  
\end{figure}

\begin{table} 
\caption{Elemental Abundances for HD 172324}                                                                 
\begin{center}
\begin{tabular}{lcccrc}
\noalign{\smallskip}                                                              
\hline       
\noalign{\smallskip}                                                              
\noalign{\smallskip}
\multicolumn{1}{l}{Species}&  
\multicolumn{1}{c}{$\log \epsilon_{\odot}$}&  
\multicolumn{1}{l}{[X/H]}&  
\multicolumn{1}{l}{s.d.}&  
\multicolumn{1}{c}{N}&  
\multicolumn{1}{r}{[X/Fe]}\\
            \noalign{\smallskip}  
            \hline  
            \noalign{\smallskip} 
  He I&10.99&$-0.08$ &$\pm$0.35 & 4 &$+0.54$\\
  He I*&10.99&$-0.19$ &$\pm$0.35 & 12 &$+0.44$\\
  C II*& 8.55 &$-1.30$ & & syn &$-0.68$\\
  O I& 8.87 & $+0.41$&$\pm$0.20&2 & $+1.03$ \\
  O I*& 8.87 & $+0.26$&$\pm$0.19&4 & $+0.89$ \\
  Ne I*& 8.09 & $+0.05$& &1 & $+0.68$ \\
  Mg I*& 7.58 &$-0.63$ & & 1  & $ 0.00$ \\
  Mg II& 7.58 &$-0.66$ &$\pm$0.17 & 2  & $-0.04$ \\
  Mg II*& 7.58 &$-0.65$ &$\pm$0.12 & 3  & $-0.02$ \\
  Al II& 6.47 &$-0.59$ & & 1&$+0.03$\\
  Al II*& 6.47 &$-0.65$ & & 1&$-0.02$\\
  Si II& 7.55 &$-0.16$ &$\pm$0.03 & 2  &$+0.47$\\
  Si II*& 7.55 &$-0.29$ &$\pm$0.14 & 6  &$+0.39$\\
  Si III& 7.55 &$+0.25$ && 1  &$+0.88$\\
  S II& 7.21 & $-0.19$&          &1 & $+0.43$ \\
  S II*& 7.21 & $-0.47$& &1 & $+0.16$ \\
  Ti II&4.98 & $-0.03$&$\pm$0.24 &  8& $+0.59$\\
  Ti II*&4.98 & $-0.28$&$\pm$0.33 &  3& $+0.35$\\
  Cr II&5.67 & $-0.43$&$\pm$0.19 & 11 & $+0.19$\\
  Cr II*&5.67 & $-0.48$&$\pm$0.28 & 3 & $+0.15$\\
Fe II&7.51 & $-0.62$&$\pm$0.17 & 17  & \\
Fe II*&7.51 & $-0.63$&$\pm$0.11 & 11  & \\
            \noalign{\smallskip}  
            \hline  
            \noalign{\smallskip}  
\end{tabular}

Notes -- same as Table 3.\

--$T_{\rm eff}$ = 11000 K, $\log g$ = 2.5 and $\xi_{t}$ = 7.00 km~s$^{-1}$ for McD spectrum\

-- $T_{\rm eff}$ = 11500 K, $\log g$ = 2.5 and $\xi_{t}$ = 5.40 km~s$^{-1}$ for OHP spectrum

* result from OHP spectrum.\\
\end{center}
\end{table}

 This star appears to be deficient in Fe by a factor of 3-4 and
 has a large radial velocity of $-$110 km~s$^{-1}$ making it a likely
 halo or old disk object.
  Carbon abundance is very important in ascertaining the evolutionary
 status. We could observe only the blend at 4267~\AA ~ while supposedly 
 strong lines at 6578 and 6582~\AA ~ were too weak to be  measured.
  By computing the blend of two C II lines at 4267~\AA ~  we find 
 [C/H] of $-$1.3 dex, for which Takeda, Takada-Hidai \& Kotake
 (1996) found  zero non-LTE correction.   In the neighbourhood of 
$T_{\rm eff}$ 10000 K the non-LTE abundance correction for carbon  
using C I lines is $\sim$ $-$0.4 dex (Venn 1995b).  Our iron abundance is based
on Fe II lines that are not strongly affected by non-LTE effects.
 For O I line at 6156-6158~\AA ~ the non-LTE correction is 
 $\sim$ $-$0.3 dex (Takeda \& Takada-Hidai 1998).

 For A-type supergiants of solar metallicity, Venn (1995a) found 
 a mean enrichment of $\sim$ 0.6 dex for Na and $\sim$ 0.3 dex for S.
 Venn (1995b) reported CNO abundances for the same sample after
 applying non-LTE corrections to the derived abundances.
 The mean values  found were: [C/H] = $-$0.4 dex, [N/H] = +0.08 dex and [O/H] = $-$0.2 dex. 
Nevertheless, the hottest star of Venn's sample, HD 161695, is about
1500~K cooler than HD 172324.

 McErlean, Lennon \& Dufton (1999) have plotted LTE and non-LTE
 line profiles for a range of temperatures 10000~K to 35000~K.
 At 11000~K the non-LTE correction for C II lines at 4267~\AA  is very
 small, in agreement with the prediction of Takeda, Takada-Hidai \& 
 Kotake (1996). Hence our values [C/H] = $-$1.3 dex and [C/Fe] = $-$0.68 dex are
realistic estimates. 

 For an A-type star of solar metallicity, a mean value [C/H] = $-$0.4 
 was reported  by Venn (1995b),
 whereas for B supergiants McErlean, Lennon \& Dufton (1999) report
 mean values of  [C/H] = $-$0.35 dex and [O/H] = $-$0.32.
 After applying non-LTE correction for oxygen we get [O/H]
  of $+$0.1 dex and [O/Fe]   of $+$0.7 dex.
 Our derived carbon and oxygen abundances in Table 6 are much different from 
 what is expected for a young B9 supergiant. 
 It has been suggested by Boothroyd \& Sackmann (1999) 
that the carbon deficiency
 can also be caused by deep circulating mixing below the base of the convective
 envelope followed by cool bottom processing (CBP) of the CNO isotopes. 
 They also showed that the CBP became more extensive at reduced 
 metallicities or at low masses. The high radial velocity for HD 172324
 (more than 100 km~s$^{-1}$) and significantly low [Fe/H] 
  makes it very likely that it has experienced CBP. 

  The only trace of an odd Z element we could observe was
 a single line of Al I, providing [Al/Fe] $\sim$ 0. One does
 not see [Al/Fe] $\sim$ +0.30 dex reported by Edvardsson et al. (1993),
 but with single line being used in our study not much significance
 could be attached to our estimated [Al/Fe].

  Hot stars at high galactic latitude were studied by Conlon et al.
 (1993a) who found very large Fe and C deficiencies in them.
  But the stars studied by these authors are much hotter than 
  HD 172324.  Neverthless, the abundance pattern of HD 172324 is
  similar to those studied by Conlon et al. (1993a;b)
  and McCausland et al. (1992).
  This star being most likely an oxygen-rich post-AGB star  
  deserves an extended analysis covering CNO and as 
significant spectral variations can be seen in four years (see Fig. 1),  a monitoring of hydrogen line
  profiles over long time scales will be of interest.

\subsection{HD 172481}

 This star shows light variations of very small amplitude, $\pm$0.15 mag
 (Van Winckel 1995).
 The spectral energy distribution  of this star
  has an unusual multi-peaked  shape (Bogart 1994).
 It was included in the study of  Van Winckel (1995)
  who, by fitting the observed flux distribution to that of Kurucz
 (1993) model atmosphere, derived $T_{\rm eff}$ = 7000~K. Van Winckel also found 
 that  the strength of emission components in hydrogen lines varied strongly.
  From the study of radial velocity variation spread
 over a decade, he reported that 
  there is not enough evidence for the binarity of the
 star. However, his spectra show considerable line splitting attributed 
 to the passage of a shock. 
  Fortunately,  we obtained  the spectrum of HD 172481 
when the atmosphere of the star was stable and therefore
 no line splitting was observed, as can be seen in the Fig. 2.   

\begin{table} 
\caption{Elemental Abundances for HD 172481}                                                                 \begin{center}
\begin{tabular}{lcccrc}
\noalign{\smallskip}                                                              \hline       
\noalign{\smallskip}                                                              
\noalign{\smallskip}
\multicolumn{1}{l}{Species}&  
\multicolumn{1}{c}{$\log \epsilon_{\odot}$}&  
\multicolumn{1}{l}{[X/H]}&  
\multicolumn{1}{l}{s.d.}&  
\multicolumn{1}{c}{N}&  
\multicolumn{1}{r}{[X/Fe]}\\
            \noalign{\smallskip}  
            \hline  
            \noalign{\smallskip} 
Li I& 1.16&$ +2.54:$& &1& $+3.16:$ \\
C I& 8.55 &$-0.62$ &$\pm$0.21 & 12 &$-0.01$\\
N I& 7.97 &$-0.63$ &$\pm$0.10 & 3 &$-0.02$\\
O I& 8.87 & $-0.58$& $\pm$0.05 &2 & $+0.04$ \\
Mg I& 7.58 &$-0.13$ &$\pm$0.21 & 2  & $+0.48$ \\
Mg II& 7.58 &$-0.06$ &$\pm$0.54 & 3  & $+0.55$ \\
Si I& 7.55 &$-0.05$ & $\pm$0.17 & 6  &$+0.57$\\
Si II& 7.55 &$-0.10$ & $\pm$0.06& 3  &$+0.52$\\
S I& 7.21 & $-0.04$&$\pm$0.16 &7 & $+0.58$ \\
K I& 5.12 & $-0.28$&  &1 & $+0.34$ \\
Ca I& 6.35 &$-0.28$& $\pm$0.19 &12 & $+0.34$\\
Sc II& 3.13 & $-0.20$& $\pm$ 0.23  &6 & $+0.41$\\
Ti I&4.98 & $-0.31$& & 1& $+0.31$\\
Ti II&4.98 & $-0.28$&$\pm$0.30 & 6& $+0.34$\\
V  II&4.00 & $-0.00$&$\pm$0.14 & 2& $+0.62$\\
Cr I&5.67 & $-0.34$&$\pm$0.05 & 3 & $+0.28$\\
Cr II&5.67 & $-0.41$&$\pm$0.16 & 10 & $+0.21$\\
Mn I &5.39 & $-0.51$&$\pm$0.11 & 3 & $+0.11$\\
Fe  I&7.51 & $-0.62$&$\pm$0.15 &  43 & \\
Fe II&7.51 & $-0.61$&$\pm$0.14 & 13  & \\
Ni I&6.25 & $-0.31$&$\pm$0.22 & 6  & $+0.30$\\
Zn I&4.60 & $-0.23$&$\pm$0.08 & 2  & $+0.39$\\
Y II& 2.23 & $+0.08$ &$\pm$0.26& 4 &$+0.69$ \\
Zr II&2.60 & $-0.14$ & & 1 &$+0.46$ \\
Ba II& 2.13 & $+0.03$ & $\pm$0.20 & 2 &$+0.65$ \\
La II& 1.21 &$-0.55$& $\pm$0.14 & 2& $+0.07$ \\
Ce II& 1.55 &$-0.24$& $\pm$0.19 & 7& $+0.38$ \\
Nd II& 1.50 &$+0.03$& $\pm$0.09 & 3& $+0.64$ \\
Eu II& 0.51 &$+0.24$&  & 1& $+0.86$ \\
            \noalign{\smallskip}  
            \hline  
            \noalign{\smallskip}  
\end{tabular}

Note -- same as Table 3.\\
\end{center}
\end{table}

 Sadly, the $S/N$ ratio of our spectrum is not very high. We could 
measure very few lines of important elements like C and O, 
hence we were not satisfied with the limited
 data  to carry out abundance analysis.
 But it was known beyond doubt that the star is highly evolved since we could 
 identify lines of C I and also saw Li I line at 6708~\AA ~. However,
 the Li I  feature was falling  at the end of our CCD frame 
 and was showing distinct doubling. One component coming nearest to Li I
 wavelength gave log $\epsilon$(Li) of  1.8. 
But $S/N$ being very poor at the end of the CCD frame, we decided
 to follow this object with  more spectra.
 At our request, David Yong of McDonald Observatory got a spectrum
 using the 2dcoude echelle spectrograph at  the 2.7m telescope of McDonald
 Observatory on May 13, 2000.  The spectrum has resolution 
 of 30,000 and wide spectral coverage  from 3900~\AA ~ to 10200~\AA ~.
 In this spectrum, we found Li I feature to be single with some asymmetry
 in the blue wing but much deeper than in July 1999.
 In all  HD 172481 spectra, the H$_{\gamma}$ has a broad absorption,
 a narrow absorption and a possible red shifted emission component.
  The H$_{\beta}$ also has a broad absorption and a narrow 
 absorption at the centre of broad absorption. There is no 
 indication of emission component.
 The H$_{\alpha}$ has a complex profile with one shallow absorption,
 one deep absorption that has emission components in both the wings.
 The blue emission component is stronger than the red one.

 The  extensive spectral  coverage of McDonald spectrum enabled us to measure a
 large number of unblended lines for several light and heavy elements. 
 As one can see from the Table 7, we could carry out a very comprehensive
 analysis of this object. 
                                     
\begin{figure}  
\centering  
\mbox{\epsfxsize=2.9in\epsfbox{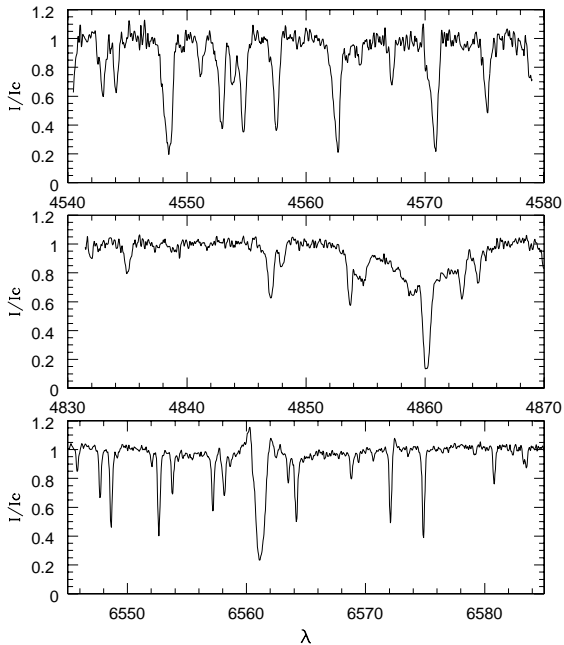}}  
\caption{Three spectral regions in the spectrum of HD 172481 clearly
showing no line splitting. Hence we believe the star was in a stable phase.}  
\end{figure}

 From Fe I and Fe II lines we derived $T_{\rm eff}$ = 7250~K and  $\log g$ = 1.5
 and was supported by Si I, Si II, Mg I and Mg II lines.                                    
 We derived the Li abundance by spectrum synthesis. All the four
 components of Li I feature at the 6708~\AA ~~ region were included.
 We find log~$\epsilon$(Li) of +3.7, which is surprisingly high.
 We were puzzled by the change in appearance of Li I feature at 6707,
 so on August 11 one more spectrum was obtained with the same set-up
 and by the same observer.
 The spectrum again showed a suggestion of doubling at the line core
 though the components were not well separated. It is not clear if the
 line is splitting periodically or the emission at the core is causing 
 it to appear double. The overall strength of Li I feature in the
August spectrum has reduced. For May 2000 spectrum,
 our  estimated Li abundance (derived by synthesis) is 
 shown in Table 7 with a colon. In the light of the large variation exhibited
 by Li I feature in strength as well as in profile shape, this value
 should be regarded with caution. However, the presence of Li I feature
 in the spectrum cannot be refuted. Another Li I feature at 6103~\AA ~~
 fell in between the echelle orders and therefore  could not be used.

 We could measure a large number of C I lines to estimate the carbon abundance.
 The three N I lines in the near infrared also enabled us to derive the
 nitrogen abundance. We get [N/Fe]$\sim$ 0 for HD 172481 whereas for
 HD 158616 [N/Fe]$= +0.3$ is reported by Van Winckel (1995). 
The estimated C/O is +0.43. 
The element Li is considered a fragile element that gets destroyed
 in the course of evolution.  The primordial abundance of Li (based
 on population II objects) is considered near log~$\epsilon$(Li) = 2.2. The excess abundance
 must therefore be caused by AGB evolution of HD 172481 or might
 owe its origin to binarity.

Our analysis covers light $s$-process elements (ls) Y and Zr 
 and heavy $s$-process elements (hs) Ba, La, Ce and Nd.
 We derive a mean [ls/Fe] of 0.6  and mean [hs/Fe] of 0.4.
 That leads to [hs/ls] of $-$0.2. With this value of [hs/ls] and [Fe/H]
 of $-$0.6 this star is remarkably similar to IRAS 04296+3429 and
 IRAS 19500-1709 studied by Van Winckel \& Reyniers (2000),
 although [ls/Fe] and [hs/Fe] are much larger for these two stars.

 The stars mentioned above and HD 158616 are very interesting post-AGB
objects as they are, most likely, evolved from intermediate-mass stars (IMS).
 According to Travaglio et al. (1999), stars in the mass range 4 - 8 M$_{\odot}$,
 activate $^{22}$Ne($\alpha$,n)$^{25}$Mg reaction during their TP-AGB
 phase more effectively than the low-mass stars due to the high 
 temperatures reached at the bottom of the convective pulse 
 (T$_{max}$ $\leq$ 3.5 $\times$ 10$^{8}$~K).
  IMS contribute more to the first $s$-peak represented
 by the elements Sr, Y and Zr. Actually, the number of known
post-AGB IMS is very small.  
 For IMS, the formation of a $^{13}$C
 pocket is less certain because of the reduced mass of the 
H-He intershell
by about one order of magnitude.  It is shown by Straniero et al.
(1997) and Gallino et al. (1998) that the $^{13}$C neutron source 
active 
during the interpulse phase of low-mass TP-AGB accounts for most 
of the
production of the second $s$-peak elements Ba, La, Ce, Sm 
and Eu.   
 
 It is obvious from the relative enhancement of Y and Sr presented 
in Tables 5 and 7 that HD 158616 and HD 172481 belong
 to the relatively rare post-AGB IMS.
 We would therefore try to understand the observed abundances of  
 HD 172481 in the framework of AGB models developed by Lattanzio and
 others for IMS.  

Lattanzio (1997) in his AGB calculation has predicted 
 that for stars more massive than 4$M_{\odot}$ the bottom of the
 convective envelope penetrates into the hotter regions of the
envelope where proton captures already modified the original CNO composition.
 This is called "hot bottom burning" (HBB) and results in
 many important changes in chemical composition of the envelope.
 Lattanzio (1997) predicted the production of Li by the
 Cameron-Fowler mechanism operating at bottom of the envelope. 
 Sackmann \& Boothroyd (1992) showed that log $\epsilon$(Li)
 $\sim$ 4.5 could be produced in stars with bolometric 
 magnitude between $-$6 and $-$7 when the temperature at the base of the
convective envelope  exceeds
 50$\times 10^{6}$~K.
 From the study of AGB stars in SMC and LMC, Smith, Plez \& Lambert
 (1995)  found them to have bolometric 
 magnitudes in the range $-$6 to $-$7.2 and to 
 show lithium values of log $\epsilon$(Li) $\sim$ 1.0 $-$ 4.0.
 However, these are cool luminous S stars.
 
 Lattanzio (1997) also predicted the destruction of carbon via CN cycle
 that could prevent C/O ratio from exceeding one.
 Theoretical models for AGB stars of mass 4, 5 and 6 M$_{\odot}$
 computed by Boothroyd, Sackmann \& Ahern (1993)  
encountered HBB with a temperature at the base of the convective
 envelope reaching 80 $\times 10^{6}$ K. These models predict C/O of 0.4 to
 0.5 for $\sim$ 10$^{3}$ yr on the AGB.

\begin{table} 
\caption{Elemental Abundances for HD 173638}                                                                 
\begin{center}
\begin{tabular}{lcccrc}
\noalign{\smallskip}                                                              
\hline       
\noalign{\smallskip}                                                              
\noalign{\smallskip}
\multicolumn{1}{l}{Species}&  
\multicolumn{1}{c}{$\log \epsilon_{\odot}$}&  
\multicolumn{1}{l}{[X/H]}&  
\multicolumn{1}{l}{s.d.}&  
\multicolumn{1}{c}{N}&  
\multicolumn{1}{r}{[X/Fe]}\\
            \noalign{\smallskip}  
            \hline  
            \noalign{\smallskip}  
C I& 8.55 &$-0.16$ &$\pm$0.13 & 5 &$-0.09$\\
Mg I& 7.58 &$-0.05$ &$\pm$0.12 & 5  & $+0.03$ \\
Si I& 7.55 &$+0.36$ &$\pm$0.12 & 2  &$+0.44$\\
Si II& 7.55 &$+0.41$ & & 1  &$+0.49$\\
Ca I& 6.35 &$+0.04$& $\pm$0.25 &12 & $+0.12$\\
Ca II& 6.35 &$-0.03$&  &1 & $+0.05$\\
Sc II& 3.13 & $+0.16$& $\pm$ 0.18  &6 & $+0.24$\\
Ti I&4.98 & $+0.01$& & 1& $+0.09$\\
Ti II&4.98 & $-0.1$&$\pm$0.18 &  17& $-0.04$\\
Cr I&5.67 & $+0.08$&$\pm$0.07 & 3 & $+0.16$\\
Cr II&5.67 & $-0.02$&$\pm$0.18 & 15 & $+0.06$\\
Mn  I&5.39 & $+0.00$&& 1 & $-0.08$\\
Fe  I&7.51 & $-0.07$&$\pm$0.13 &  62 & \\
Fe II&7.51 & $-0.08$&$\pm$0.14 & 19  & \\
Ni I& 6.25 & $+0.06$ & $\pm$0.17&7 &$+0.14$ \\
Zn I& 4.60 & $-0.06$ &$\pm$0.05  & 2 &$+0.02$ \\
Y II& 2.23 & $+0.02$ &$\pm$0.22& 5 &$+0.10$ \\
Zr II&2.60 & $-0.02$ & & 1 & $+0.06$ \\
Ba II& 2.13 & $+0.15$ & $\pm$0.04 & 2 &$+0.23$ \\
Ce II& 1.58 & $+0.02$&&1&$+0.10$ \\
Nd II& 1.48 & $+0.11$&&1& $+0.19$\\
            \noalign{\smallskip}  
            \hline  
            \noalign{\smallskip}  
\end{tabular}

Note -- same as Table 3.\\
\end{center}
\end{table} 

 The abundance pattern of  HD 172481 bears some resemblance to those
 found for HR 7671 though the latter is more metal-poor. 
 Interestingly, HR 7671 also shows the Li I feature though Li
  is not as overabundant as in  HD 172481.
  The estimated C/O for HR 7671 is 0.4. The mean [$\alpha$/Fe] is 
  lesser than what we find for HD 172481 but [s/Fe] is comparable.

 The existence of objects like HD 172481 and HR 7671 lend further support
 to the HBB scenario, put forward to explain the paucity
  of carbon-rich stars among AGB stars.
 The observed  variation of Li I feature in HD 172481 makes it
 a  very promising candidate for binarity search.

While the present paper was under the refereeing process, the referee
called our attention to the 
 then unpublished work on HD 172481 by Reyniers \& Van Winckel 
(2001). Thus a comparison with their results is most appropriate. 
These authors have derived atmospheric parameters ($T_{\rm eff}$= 7250 K and
$\log g$ = 1.5)   that are in excellent agreement with those derived by us.
 A small difference in $\xi_{t}$ of 0.6 km s$^{-1}$ is seen. These authors
 also find a large lithium abundance similar to our finding. The [Fe/H]
 and other Fe-peak elements have very good agreement. The $s$-process
 elements La, Nd and Eu however, show some disagreement. 
We have measured 3 lines 
 of Nd II so we are surprised by the lower limits placed by these authors.
 For Eu, we measured a fairly clean line at 6645~\AA~ whereas Reyniers
 \& Van Winckel used spectrum synthesis. It should be noted that the 
 oscillator strengths used by these authors are quite different from 
 those employed by us, that can possibly explain the abundance differences
 for the $s$-process elements. The $gf$ values for the lines  of these 
 elements are known to
 have larger uncertainties compared to those of the Fe-peak elements.
 Another important finding by Reyniers \& Van Winckel is the detection
 of a red luminuous companion, found from the presence of TiO bands in the 
 red and also from the observed spectral energy distribution. 
 As mentioned above, 
the presence of the companion might explain the Li I feature variations
observed by us.

\subsection{HD 173638}

This star (HR 7055) was observed in the 13-Colour photometric system and its
temperature was calculated by various approaches by Bravo Alfaro, Arellano Ferro
\& Schuster (1997).
The  photometric temperature  estimates range between 7486~K and 8100~K.
 The spectrum being of $S/N$ = 95, we could measure a large 
 number of lines covering many important 
 elements like C, $\alpha$-process elements and Fe-peak elements. 
 From the ionisation equilibrium of Fe~I$/$Fe~II, Cr I$/$Cr II,
 Ti I$/$Ti II, Si I$/$Si II and even Ca~I$/$Ca~II the
 atmospheric parameters are well-determined and listed in Table 2.

 The star appears to have
 near-solar abundances for most elements except Si which is overabundant by a factor of 3 (Table 8).
 With its small radial velocity (+11 km~s$^{-1}$), it is most likely 
 a young  star belonging  to  the disk population.
  But it can serve
 as excellent calibrator of photometric indices in the 7500 K
 temperature range. 

\subsection{HD 218753}

This star was classified as  A5III by Cowley et al. (1969). These and
the Sp.T. - Colour - $T_{\rm eff}$ calibration of Schmidt-Kaler (1982) give a temperature of 8091 K.

\begin{figure}  
\centering  
\mbox{\epsfxsize=2.9in\epsfbox{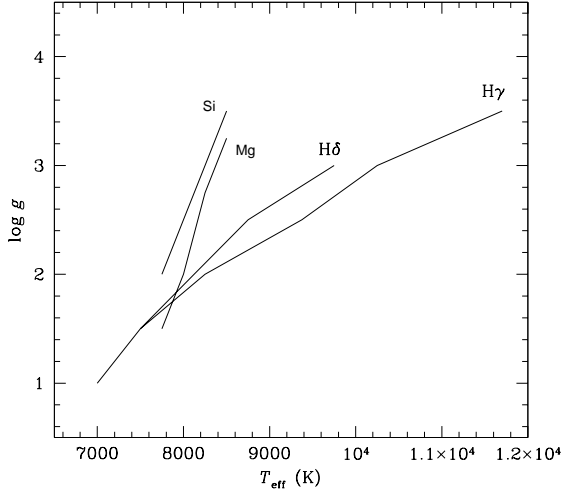}}  
\caption{H$_\gamma$, H$_\delta$, Mg and Si loci for HD 218753.}  
\end{figure}

\begin{table} 
\caption{Elemental Abundances for HD 218753}                                                                 
\begin{center}
\begin{tabular}{lcccrc}
\noalign{\smallskip}                                                              
\hline       
\noalign{\smallskip}                                                              
\noalign{\smallskip}
\multicolumn{1}{l}{Species}&  
\multicolumn{1}{c}{$\log \epsilon_{\odot}$}&  
\multicolumn{1}{l}{[X/H]}&  
\multicolumn{1}{l}{s.d.}&  
\multicolumn{1}{c}{N}&  
\multicolumn{1}{r}{[X/Fe]}\\
            \noalign{\smallskip}  
            \hline  
            \noalign{\smallskip}  
C I& 8.55 &$-0.49$ &$\pm$0.18 & 5 &$-0.30$\\
O I& 8.87 & $-0.14$&   &1 & $+0.05$ \\
Na I& 6.32 & $+0.32$&$\pm$0.07 &2 &$+0.51$  \\
Mg I& 7.58 &$-0.27$ &$\pm$0.16 & 5  & $-0.08$ \\
Mg II& 7.58 &$-0.29$ &$\pm$0.15 & 4  & $-0.10$ \\
Si I& 7.55 &$+0.11$ &$\pm$0.15 & 3  &$+0.20$\\
Si II& 7.55 &$+0.00$ &$\pm$0.22 & 4  &$+0.26$\\
S I& 7.21 & $+0.17$&$\pm$0.09 &2 & $+0.36$ \\
Ca I& 6.35 &$-0.08$& $\pm$0.15 &15 & $+0.11$\\
Ca II& 6.35 &$-0.09$&  &1 & $+0.10$\\
Sc II& 3.13 & $+0.00$& $\pm$ 0.27  &8 & $-0.19$\\
Ti I&4.98 & $-0.27$& & 1& $-0.08$\\
Ti II&4.98 & $-0.25$&$\pm$0.17 &  19& $-0.06$\\
Cr I&5.67 & $-0.01$&$\pm$0.38 & 3 & $+0.18$\\
Cr II&5.67 & $-0.16$&$\pm$0.13 & 14 & $+0.03$\\
Mn  I&5.39 & $-0.23$&$\pm$0.06 & 3 & $-0.04$\\
Fe  I&7.51 & $-0.20$&$\pm$0.16 &  82 & \\
Fe II&7.51 & $-0.18$&$\pm$0.13 & 30  & \\
Ni I& 6.25 & $-0.06$ & $\pm$0.15&9 &$+0.13$ \\
Zn I& 4.60 & $+0.18$ & & 1 &$+0.37$ \\
Y II& 2.23 & $-0.24$ &$\pm$0.21& 7 &$-0.05$ \\
Zr II&2.60 & $+0.18$ & & 1 & $+0.37$ \\
Ba II& 2.13 & $-0.08$ & $\pm$0.16 & 2 &$+0.11$ \\
            \noalign{\smallskip}  
            \hline  
            \noalign{\smallskip}  
\end{tabular}

Notes -- same as Table 3.\\
\end{center}
\end{table}

Photometric data in the Str\"omgren's system can also be used to estimate $T_{\rm eff}$, through the $\beta -(b-y)$ relation
of Crawford (1979) (his Table 1) and adopting $(b-y) = 0.236$ and 
$\beta = 2.773$ (Hauck \& Mermilliod 1998).
This leads to $(b-y)_o$ = 0.176 and $E(b-y) = 0.054$. While $(b-y)_o$ leads
to $(B-V)_o$ = 0.20-0.25 (Crawford 1970), this implies $T_{\rm eff}$ = 7800 K.

Also, if the above colour excess and photometry are used in combination of 
Napiwotzki, Sh\"onberner \& Wenske's (1993) calibration we 
find $T_{\rm eff}$ = 7200 K. 
However, it must be emphasized that
the calibration has been computed using only stars of luminosity classes
V and IV, while HD 218753  could be of  luminosity class III or II.

 Balmer line fitting can also be used to estimate $T_{\rm eff}$. Theoretical profiles have been extracted
from Kurucz's (1993) models for H$_\alpha$, H$_\beta$, H$_\gamma$ and H$_\delta$. The observed 
profiles were fitted with theoretical profiles of given $T_{\rm eff}$ and $\log g$.
The solution is not unique but in fact the best fits define a locus on the
$T_{\rm eff}$ - $\log g$ plane for each Balmer line. As underlying core emission may be
present in some stars, we chose  to fit the wings H$_\gamma$ and H$_\delta$. 
Following the above fitting process we found the loci for H$_\gamma$ and H$_\delta$ profiles
shown in Fig. 3.

In hot stars, magnesium and silicon lines can be used as $T_{\rm eff}$ 
indicators (since 
 Fe I lines are strongly affected by non-LTE effects).
 Given a pair ($T_{\rm eff}$, $\log g$) one searches for abundance 
consistency between 
 neutral and ionized lines of  Mg and Si. 
Again, the solution is not unique 
but rather a locus on the $T_{\rm eff}$ - $\log g$ plane is defined for each element, as illustrated in
Fig. 3. The intersections of the Mg and  Si loci with the H$_\gamma$ and H$_\delta$ loci
point to the proper temperature and gravity. In this fashion 
we estimated $T_{\rm eff}$ between
7600 and 7900 K and $\log g$ between 1.75 and 2.0.

The turbulent velocity, $\xi_{t}$, was estimated from  Fe II lines by requiring abundance to be independent of line strength. We found $\xi_{t}$ = 3.3 km~s$^{-1}$.

\begin{table} 
\caption{Elemental Abundances for HD 331319}                                                                 
\begin{center}
\begin{tabular}{lcccrc}
\noalign{\smallskip}                                                              
\hline       
\noalign{\smallskip}                                                              
\noalign{\smallskip}
\multicolumn{1}{l}{Species}&  
\multicolumn{1}{c}{$\log \epsilon_{\odot}$}&  
\multicolumn{1}{l}{[X/H]}&  
\multicolumn{1}{l}{s.d.}&  
\multicolumn{1}{c}{N}&  
\multicolumn{1}{r}{[X/Fe]}\\
            \noalign{\smallskip}  
            \hline  
            \noalign{\smallskip}  
C I& 8.55 &$-0.33$ &$\pm$0.30 & 5 &$-0.09$\\
O I& 8.87 & $+0.06$&$\pm$0.02   &2 & $+0.30$ \\
Na I& 6.32 & $-0.03$&&1 &$+0.21$  \\
Mg I& 7.58 &$-0.32$ &$\pm$0.21 & 3  & $-0.08$ \\
Mg II& 7.58 &$-0.18$ &$\pm$0.12 & 3  & $+0.06$ \\
Al I& 6.47 &$+0.07$ & & 1&$+0.31$\\
Si II& 7.55 &$-0.18$ &$\pm$0.20 & 2  &$+0.06$\\
S I& 7.21 & $+0.29$&$\pm$0.18 &5 & $+0.53$ \\
Ca I& 6.35 &$-0.29$& $\pm$0.18 &10 & $-0.05$\\
Sc II& 3.13 & $-0.19$& $\pm$ 0.22  &5 & $+0.05$\\
Ti II&4.98 & $-0.33$&$\pm$0.22 &  22& $-0.09$\\
V II&4.01 &$+0.19 $& $\pm$0.16 & 3 &+0.43 \\
Cr I&5.67 & $-0.01$&$\pm$0.29 & 3 & $+0.23$\\
Cr II&5.67 & $+0.02$&$\pm$0.14 & 18 & $+0.22$\\
Mn  I&5.39 & $-0.17$& & 1 & $+0.07$\\
Fe  I&7.51 & $-0.27$&$\pm$0.16 &  48 & \\
Fe II&7.51 & $-0.20$&$\pm$0.15 & 21  & \\
Ni I& 6.25 & $+0.00$ & $\pm$0.22&4 &$-0.24$ \\
Ni II& 6.25 & $-0.13$ & &1 &$+0.11$ \\
Sr II& 2.90 &$+0.01$& &1&$+0.25$ \\
Y II& 2.23 & $-0.58$ &$\pm$0.09& 3 &$-0.34$ \\
Ba II& 2.13 & $-0.51$ & $\pm$0.16 & 3 &$-0.27$ \\
            \noalign{\smallskip}  
            \hline  
            \noalign{\smallskip}  
\end{tabular}

Notes -- same as Table 3.\\

\end{center}
\end{table} 

\begin{table*}[!]
\caption{Absortion lines in the spectrum of the proto-Planetary Nebula HDE 341617}
\begin{center}
\begin{tabular}{lccrc|lccrc}
\noalign{\smallskip}                                                              \hline
\noalign{\smallskip}
\multicolumn{1}{l}{Species}&
\multicolumn{1}{c}{$\lambda_{obs}$}&
\multicolumn{1}{c}{$\lambda_{lab}$}&
\multicolumn{1}{l}{$Eq. W.$}&
\multicolumn{1}{c}{$Vr$}&
\multicolumn{1}{l}{Species}&
\multicolumn{1}{c}{$\lambda_{obs}$}&
\multicolumn{1}{c}{$\lambda_{lab}$}&
\multicolumn{1}{l}{$Eq. W.$}&
\multicolumn{1}{c}{$Vr$}\\
\multicolumn{1}{l}{}&
\multicolumn{1}{c}{(~\AA ~)}&
\multicolumn{1}{c}{(~\AA ~)}&
\multicolumn{1}{l}{($m$~\AA )}&
\multicolumn{1}{l}{$(km~s^{-1})$}&
\multicolumn{1}{l}{}&
\multicolumn{1}{c}{(~\AA ~)}&
\multicolumn{1}{c}{(~\AA ~)}&
\multicolumn{1}{l}{($m$~\AA )}&
\multicolumn{1}{c}{$(km~s^{-1})$}\\
            \noalign{\smallskip}
            \hline
            \noalign{\smallskip}

He I&4010.217&4009.270&223.2&70.9&O II&4642.973&4641.811 &233.6 &75.1\\

He I&4027.047&4026.362&337.0&48.6&O II&4643.039&4641.82&302.9&78.8\\

He I&4144.569&4143.759&391.9&51.0&O II&4650.29&4649.139&309.0&74.3\\
O II:&4350.53&4349.426&267.7&76.1&O II&4662.795&4661.635&150.4&74.7\\

O II&4367.922&4366.896&208.3&70.5&O II+C III&4674.776&4673.91/73.75&58.6& \\

He I&4389.027&4387.929&325.8&75.1&O II&4677.389&4676.234&84.8&74.1\\

O II&4415.984&4414.909&141.5&73.0&O II&4706.507&4705.355&70.2&73.4\\

O II&4418.081&4416.975&104.7&75.1&He I&4714.39&4713.143&160.8&79.4\\

He I&4472.59&4471.477&446.8&74.7&He I&4922.992&4921.929&516.4&64.8\\

Si III&4553.727&4552.654&322.4&70.7&He I:&5015.647&5015.675&217.2& \\
Si III&4568.918&4567.872&280.0&68.7&Fe II&5796.901&5795.870&152.8&53.4\\
Si III&4575.818&4574.777&229.7&68.3&He Ibld&5875.849&5875.618+75.650&542.5&\\
O II& 4592.136&4590.971&209.9&76.1&He I&6679.176&6678.149&841.9&46.1\\

O II& 4639.951& 4638.854&136.3&70.9&&&&&\\
\noalign{\smallskip}
            \hline
            \noalign{\smallskip}
\end{tabular}

\end{center}
\end{table*}

For HD 218753 we  finally adopted $T_{\rm eff}$ = 8000 K, $\log g$ = 2.0 dex and 
$\xi_{t}$ = 3.3  km~s$^{-1}$. With these
parameters the abundances for the rest of the detected species were 
calculated and the  results are given in Table 9.
 
 For this star our spectrum has $S/N$ = 67 and we could
 measure a large number of clean weak lines for most important elements.
  HD 218753 shows a significant carbon deficiency most likely caused by
 CNO processing. There is also a significant enrichment of sodium
 that has been found in  many A-F supergiants as discussed before for HD 725 (Takeda and Takada-Hidai 1994).  
  Among $\alpha$-capture elements, only S shows enrichment above
 detection limit. These are indications of the star having
 experienced the first dredge-up (e.g. [C/Fe] = $-$0.3 dex).
 However its position in the H-R diagram of Fig. 5 is consistent with
a $M\sim1.5-2 M_{\odot}$ and an age of about $7.9 \times 10^{8}$ yr.

\subsection{HD 331319}

This star  is an IRAS source (19475+3119). 
 The IR fluxes are quite large (Table 1), the heliocentric radial velocity is very small
($-$2.5 km~s$^{-1}$). With a spectrum of $S/N$ = 98 we could measure a  large number of
 lines and hence derive the atmospheric parameters as well as abundances
 with good accuracy (Table 10). The star is Fe-poor by about 1.8 times, but  
shows significant enrichment of sulphur. The derived abundances are
 based on 5 good lines hence the derived sulphur abundance cannot be 
 ascribed to measurement errors. Similarly, vanadium also shows some enrichment.
 Sulphur enrichment appears to be a common feature of our sample stars.
 The derived abundance of carbon for HD 331319  
clearly indicates the effect of CN  processing. The star is most likely
a young massive disk supergiant or bright giant
 ascending the red giant branch. 
 The IR fluxes are probably caused by material ejected at this evolutionary stage.

\subsection {The proto-Planetary Nebula HDE 341617}

\begin{figure}[!]
\centering
\mbox{\epsfxsize=2.9in\epsfbox{f4.epsi}}
\caption{A comparison of equivalent widths of the emission spectrum of HDE 341617 as observed in 1993 and in 1999. (a) The weakening of the emission spectrum is evident despite the fact that the 1999
spectra have not been flux calibrated. (b) The equivalent widths are not a function of wavelength. However, the large scatter is probably due to the lack of calibration. See text for discussion.}
\end{figure}

\begin{table*}[!]
\caption{Emission lines in the spectrum of the proto-Planetary Nebula HDE 341617}
\begin{center}
\begin{tabular}{lccrc|lccrc}
\noalign{\smallskip}
\hline
\noalign{\smallskip}                                                              \noalign{\smallskip}
\multicolumn{1}{l}{Species}&
\multicolumn{1}{c}{$\lambda_{obs}$}&
\multicolumn{1}{c}{$\lambda_{lab}$}&
\multicolumn{1}{l}{$Eq. W.$}&
\multicolumn{1}{c}{$Vr$}&
\multicolumn{1}{l}{Species}&
\multicolumn{1}{c}{$\lambda_{obs}$}&
\multicolumn{1}{c}{$\lambda_{lab}$}&
\multicolumn{1}{l}{$Eq. W.$}&
\multicolumn{1}{c}{$Vr$}\\
\multicolumn{1}{l}{}&
\multicolumn{1}{c}{(~\AA ~)}&
\multicolumn{1}{c}{(~\AA ~)}&
\multicolumn{1}{l}{($m$~\AA )}&
\multicolumn{1}{l}{$(km~s^{-1})$}&
\multicolumn{1}{l}{}&
\multicolumn{1}{c}{(~\AA ~)}&
\multicolumn{1}{c}{(~\AA ~)}&
\multicolumn{1}{l}{($m$~\AA )}&
\multicolumn{1}{l}{$(km~s^{-1})$}\\
            \noalign{\smallskip}
            \hline
            \noalign{\smallskip}

[S II]&4069.357&4068.600&545.7&55.8&[Cr I: ]&5147.604&5146.550& 86.5&61.4\\

[Fe II]&4244.732&4243.980&266.3&53.1&[Fe II]&5158.931&5158.000&155.2&54.1\\

[Fe II]&4245.566&4244.810&50.0&53.4&[Fe II]&5159.702&5158.810&367.9&51.8\\

[Fe II]&4277.632&4276.83&237.5&56.2&[Fe II]&5164.910&5163.940&108.9&56.4\\

[Fe II]&4306.718&4305.900&60.2&57.0& Fe II&5169.933&5169.030&164.1&52.4\\

H$_\gamma$&4341.315&4340.475&1030.0& &[Fe II]&5182.956&5181.970&89.0&57.1\\

[Fe II]&4353.553&4352.78& 100.5&58.8&[Fe II]&5200.056&5199.180&43.0&50.5\\

[Fe II]&4359.146&4358.370&117.9&53.4&[Fe II]&5221.008&5220.060&74.3&54.5\\

[Fe II]&4360.112&4359.190&496.6&53.1&[Fe II]&5262.580&5261.610&463.8&55.3\\

O I&4369.018&4368.30&100.9&49.3&[Fe II]&5269.799&5268.880&58.5&52.3\\

Fe II&4373.130&4372.220&37.1&62.4&[Fe II]&4288.157&4287.40&625.2&53.0\\

[Fe II]&4383.621&4382.750&81.0&59.6&[Fe II]&5274.345&5273.380&383.7&54.9\\
Fe II& 4385.381&4384.330&24.7&71.9&O I&5276.020&5275.080&29.2&53.5\\

[Fe II]&4414.564&4413.780&398.2&53.3&Fe II&5279.221&5278.265&21.1&54.3\\

[Fe II]&4417.075&4416.270&356.9&54.7&Fe III&5292.588&5291.780&46.7&45.0\\

[FeII]&4458.737&4457.95&221&52.9&[Fe II]&5297.799&5296.840&78.1&54.3\\

[Fe II]&5413.604&5412.640&50.3&53.4&O I&5299.961&5299.000&117.2&54.4\\

[FeII]&4475.767&4474.91& 85&53.0 &Fe II&5317.554&5316.609&23.9&53.3\\

[FeII]&4489.560&4488.75& 85&51.0&[Fe II]&5334.587&5333.650&295.0&52.7\\

[FeII]&4493.412&4492.64& 38&53.4&[Fe II]&5348.652&5347.690&30.0&54.0 \\

[FeII]&4529.249&4528.39& 18&56.4&[Fe II]&5377.438&5376.470&234.3&54.0\\

[FeII]&4728.955&4728.07&170&56.4&[Fe II]&5434.124&5433.150&86.1&53.8\\

Fe III&4734.765&4733.900&14.9&54.8&[Fe II]&5478.195&5477.250&43.9&51.8\\

[FeII]&4775.615&4774.74&110&54.4&O I&5513.699&5512.710&30.0&53.8\\

[Fe II]&4799.149&4798.29&25.9&53.7&O I&5555.937&5554.940&60.0&53.8\\

[Fe II]&4815.394&4814.5&325.9&52.8&[Fe II]&5747.997&5746.960&130.2&54.1\\

H$_\beta$&4862.284&4861.332&2921.0&58.7&[N II]&5755.637&5754.800&157.5&43.6\\

[Fe II]&4875.354&4874.490&116.5&53.2&Si II&5958.641&5957.612&149.5&51.8\\

[Fe II]&4890.511&4889.630&288.0& 54.1&O I&5959.607&5958.46+58.630&&\\

[Fe II]&4906.204&4905.350&256.0& 52.2&Si II o FeIII&5979.989&5978.970&425.3&51.1\\
Fe II&4924.813&4923.921&86.3& 54.3&O I bld&6047.481& 6046.26+6046.46&427.2&60.6\\

[Fe II]&4951.656&4950.740& 46.0& 55.5&Si II(2)+[NII]&6348.259&6347.090&517.8&55.3\\

[Fe II]&4974.283&4973.390&101.5&53.9&[O I]&6364.896&6363.88&277.7&47.9\\
uf$^1$&4981.034&&113.5&&uf$^1$&6366.259&&126.3&\\

[Fe II]&5006.441&5005.520&69.1&43.2&Si II(2)&6372.505&6371.359&242.9&54.0\\
Fe II&5019.322&5018.434&97.5&53.17&[N II](1)&6549.274&6548.100&666.1&53.8\\

[Fe II]&5021.110&5020.240&107.4&52.0&H${\alpha}$& 6564.138&6562.817&43.97&60.3\\
Si II&5041.962&5041.063&61.9&53.5&[N II](1)&6584.646&6583.6&2105.0&47.7\\

[Fe II]&5044.422&5043.530&58.1&53.1&[O II]&6668.000&6666.940&160.2&47.\\
Si II&5056.882&5056.020&178.0&51.1&[S II]&6717.786&6716.470&81.1&58.8\\

[Fe II]&5108.814&5107.950&57.1&50.7&[S II]&6731.874&6730.850&93.3&45.6\\

[Fe II]&5112.558&5111.630& 88.8&54.5&[S II]&6732.169&6731.300&114.5&38.72\\
\noalign{\smallskip}
            \hline
            \noalign{\smallskip}
\end{tabular}

1 -- uf = unidentified feature.\\
\end{center}
\end{table*}

 It is an interesting object that has been showing relatively
 fast changes in temperature and brightness. It has faded from
 m$_{v}$ = 8.8 reported in BD catalogue to m$_{v}$ = 11.4
 estimated by Stephenson (1986). Extensive photometry was taken up
 by Arkhipova et al. (1999), who found rapid light variations with an
 amplitude of up to 0.3 mag during 1996-1997. These authors also obtained
 a spectrum and describe the absorption and emission lines present.
 The light variations of HDE 341617 and spectral appearance
 confirmed the suggestion of Volk \& Kwok (1989), based on IRAS 
 color indices, that it is a candidate post-AGB star.
 From spectral type A5 given in HDE, it had changed to class Be (1986)
 as reported by Downes \& Keyes (1988).
 A very extensive spectral investigation by Parthasarathy et al. (2000)
 has shown the star to have $T_{\rm eff}$ = 22000 K with large deficiency
 of carbon. These authors derived $T_e$ of 10000 K and $N_e$
 of 2.5 $\times$ 10$^{4}$ cm$^{-3}$ for the nebula from the study of emission lines.
 The spectra used by Parthasarathy et al. were taken
 in 1993. As this object presents rapid variations, we felt we could
 examine the spectrum at our epoch (1999) and study the changes. 
A comparison of the emission features common to 
Parthasarathy et al. 
(2000) 1993 spectrum and ours from 1999 is presented in Fig. 4. Fig. 4a 
shows that in 1999 the emission lines were weaker while Fig. 4b shows
that the differences in equivalent widths are not a function of wavelength.
Unfortunately, we did not observe any flux standard and so were unable to
measure the fluxes     
 in the emission lines. The shape of the pseudo continuum used for each
 echelle order may also be affected by instrumental sensitivity
 function. The large scatter is due to this fact.  Weakening of lines (if it is not caused by resolution difference)
 might indicate further increase in temperature of the central star.
 Systematic monitoring of this fast evolving object could be very rewarding.

According to Arkhipova et al. (1999) HDE 341617 
has a  mass of 0.7 $M_{\odot}$  with 
an envelope of $\sim$ 10$^{-3}$ $M_{\odot}$ undergoing rapid evolution towards 
the PN phase which, according to the 
predictions from Bl\"ocker's (1995) models, should be
reached within 100 years. This and the rapid photometric variations reported 
by Arkhipova et al. (1999),  most likely caused 
by variations in the stellar wind,
make the  star a very interesting target for continuous monitoring. 

  We have two relatively low $S/N$ spectra that were used for  identification and measurement of absortion and emission lines. We      
present in Tables 11 and 12  the line strengths and 
velocities for absorption and emission lines for this object.

 Several absortion lines of  He I, O II, Si III and Fe II are 
detected. The  radial velocity, for each individual line 
unambiguously identified is given in Table 11. For He I  two groups of 
lines are identified at average velocities of 52.6$\pm$8.3 km~s$^{-1}$
(4 lines) and 75.0 $\pm$ 3.7 km~s$^{-1}$ (4 lines). This suggests
some stratification in the atmosphere. The rest of the species however average 
73.3$\pm$2.9 km~s$^{-1}$ (15 lines). The low  dispersion in the radial 
velocity suggests that all these lines are formed in the same region of the stellar atmosphere and no stratification is evident.

The emission spectrum, formed in the outer envelope and/or in the nebulosity, 
 consists of lines of  Fe II, [Fe II], 
Fe III, [S II], O I, [N II] and Si II.
 The radial velocities of emission lines are all very 
consistent and average 54.3$\pm$3.4 km~s$^{-1}$.
 The difference of radial velocities between the 
absorption and emission lines indicates that the 
nebulosity expands at about 19 km~s$^{-1}$.
The Balmer lines show emission  on top of the stellar absorption.
The emission is shifted relative to the absorption and 
is consistent with the radial velocities of other emission features, 
showing that it is also produced in the same region.

 Though the number of absorption lines were relatively small, we have 
 done an abundance analysis for a few elements.
 The O II lines were used for fixing the microturbulence velocity.
 Since the line data was not adequate to do a detailed excitation
 equilibrium, we chose to use the temperature and gravity estimated
 by Parthasarathy et al. (2000) as starting value and tried models both
 hotter and cooler than their estimate. We got more consistent
 values for $T_{\rm eff}$ = 23000 K, $\log g$ = 3.0 dex and $\xi_t$ = 15.0 km~s$^{-1}$ and hence these
 parameters were adopted although available lines were not particularly sensitive  
 to the temperature. Temperature errors of $\pm$500~K or more 
 are possible. 
With these parameters the atmospheric chemical abundances 
 for the central star
are those reported in Table 13.

\begin{table}
\caption{Elemental Abundances for the central star of the PPN HDE 341617.}
\begin{center}
\begin{tabular}{lcccr}
\noalign{\smallskip}
\hline
\noalign{\smallskip}
\noalign{\smallskip}
\multicolumn{1}{l}{Species}&
\multicolumn{1}{c}{$\log \epsilon_{\odot}$}&
\multicolumn{1}{l}{[X/H]}&
\multicolumn{1}{l}{s.d.}&
\multicolumn{1}{c}{N}\\
            \noalign{\smallskip}
            \hline
            \noalign{\smallskip}
C II& 8.55 &$-1.43$ & & 1 \\
N II& 7.97 &$-0.50$ & $\pm$0.30&3\\
O II& 8.87 & $-0.51$&$\pm$0.21&14 \\
Mg II& 7.58 &$-1.13$& & 1\\
Si III& 7.55 &$-0.63$ &$\pm$0.06 & 2\\
            \noalign{\smallskip}
            \hline
            \noalign{\smallskip}
\end{tabular}

Notes -- same as Table 3.\\

\end{center}
\end{table}

\section {Discussion}

In order to estimate the mass and the age of each of the program stars
we have plotted them on the H-R diagram as can be seen in Fig. 5.
The effective temperatures are those derived from the chemical analysis,
and the luminosities were obtained from the above
temperatures and the calibration of Schmidt-Kaler (1982); these quantities
are listed in Table 2. Despite
parallaxes do exist in the Hipparchos catalogue for some of the stars in
our sample, we have
preferred the above approach for the luminosity calculation 
 for the following reasons. While a very respected version of the
 Period - Luminosity relation
for cepheids has been calculated using Hipparchos parallaxes 
(Feast \& Catchpole 1997) using the most well-known  26 cepheids 
as calibrators,
individual parallaxes seem to be of little use. We have taken 21 cepheids 
such that the distances could be estimated without requiring the use
 of parallaxes. A straight comparison of 
 distance moduli obtained from the Hipparchos parallaxes and  from the 
P-L relation shows, in many cases,
large differences. We have also noticed that Hipparchos parallaxes place some stars
at unacceptable positions on the H-R diagram given the  derived atmospheric
 parameters and elemental
 abundances.  
 For example, for stars 
 HD 172324 and HD 172481  the parallax 
data suggest values of log $(L/L_{\odot}) < 2.7$. The situation is further 
 complicated by  unknown circumstellar reddening, which could be  
substantial for some of the stars, given their high infrared fluxes.
Therefore we decided not to use individual parallaxes in the present context.

All pre-AGB evolutionary model  tracks plotted  
in Fig. 5 are those of Schaller et al.
(1992), while the post-AGB models are from Bl\"ocker (1995). All isochrones
are from the work of Bertelli et al. (1994).

\begin{table*}
\caption{Comparison of relevant [X/Fe] ratios for program stars and well known post-AGB stars.}
\begin{center}
\begin{tabular}{lcccccccccc}
\noalign{\smallskip}
\hline
\noalign{\smallskip}
\noalign{\smallskip}
\multicolumn{1}{c}{Star}&
\multicolumn{1}{c}{[Fe/H]}&
\multicolumn{1}{c}{[C/Fe]}&
\multicolumn{1}{c}{[O/Fe] }&
\multicolumn{1}{c}{[Mg/Fe] }&
\multicolumn{1}{c}{[Si/Fe]}&
\multicolumn{1}{c}{[S/Fe]}&
\multicolumn{1}{c}{[Ca/Fe]}&
\multicolumn{1}{c}{[s/Fe]}&
\multicolumn{1}{c}{C/O}&
\multicolumn{1}{c}{reference}\\
            \noalign{\smallskip}
            \hline
            \noalign{\smallskip}

HD 725& $-0.29$& $-0.08$ & &$+0.12$&$+0.40$&+0.20&+0.14&+0.27& &1\\
HD 9167&$-0.34$ & & &$-0.04$ & &&$+0.12$& $+0.12$ &&1 \\
HD 172324&$-0.63$& $-0.68$& +0.96&$-0.02$&$+0.43$&$+0.42$ && &+0.01 &1\\
HD 173638&$-0.08$&$-0.09$& &+0.03&$+0.46$& &$+0.09$& $+0.14$& &1\\
HD 218753&$-0.19$&$-0.30$&+0.05&$-0.09$&+0.23&+0.36&$+0.11$&$+0.14$&+0.21 &1\\
HD 331319&$-0.24$&$-0.09$&+0.30&$-0.01$&+0.06&$+0.53$&$-0.05$&$-0.30$&+0.19 &1\\
            \hline
HD 158616&$-0.57$ &+0.34&+0.04 &+0.08&$+0.56$ &+0.62&+0.21& +0.66& $+0.95$&1\\
 HD 158616&$-0.70$ &+0.64&$-0.04$ &&+0.68&$+0.70$&+0.71&+1.30&$+1.90$&2\\
HD 172481&$-0.61$&-0.01&+0.04&+0.51&+0.54&$+0.58$ &$+0.34$&+0.52&$+0.43$ &1\\
HR 7671&$-1.10$&$-0.40$&$-0.30$&+0.25&$+0.40$&$+0.15$&+0.32&$+0.60$&$+0.38$& 6\\
            \hline
HD 187785&$-0.40$&$+1.00$&$+0.60$&+0.67&$+0.82$&+0.57&$+0.49$&$+1.30$&$+1.20$&4\\
HD 187785 &$-0.60$&$+1.00$&$+0.70$&+0.38& &+0.26&$+0.49$&$+1.10$& &11\\
HD 56126&$-1.00$&$+1.08$ &$+0.63$&$+0.97$&$+0.95$&$+0.63$&$+0.46$&$+1.78$ &$+1.35 $&3\\
HD 56126&$-1.00$&$+1.10$ &$+0.80$&$+0.06$& &$+0.40$&$-0.11$&$+1.50$ & &11\\
IRAS04296$+$3429&$-0.60$&$+0.80$& & &$+0.79$&$+0.43$&$+0.18$&$+1.50$& &11\\
IRAS05341$+$0852&$-0.80$&$+1.00$&$+0.60$& &$+0.59$&$+0.28$&$+0.08$&$+2.20$& &11\\
IRAS22223$+$4237&$-0.30$&$+0.30$&$-0.10$& &$+0.29$&$+0.04$&$-0.17$&$+0.90$& &11\\
IRAS23304$+$6147&$-0.80$&$+0.90$&$+0.20$& &$+0.79$&$+0.56$&$+0.29$&$+1.60$& &11\\
            \hline
HR 4049&$<-3.2$&$>+3.0$&$>+2.7$&$>+1.7$&$>+0.40$&$>+3.0$ &$<-2.1$& &+0.95 &5\\
 HD 44179&$-3.30$&$+3.30$&$+2.90$&+1.2 &$+1.50$&$+3.00$& &+0.20 &$+1.20$&2\\
 HD 46703&$-1.57$&$+0.98$&$+1.10$&$+0.09$&$-0.38$ &$+1.20$&$+0.02$&$-0.49$&$+0.74$&7,8\\
 HD 52961&$-4.80$&$+4.40$&$+4.20$& &&$+3.80$& &$+0.60$&$+0.76$&2\\
HD 70379&$-0.31$&$+0.42$ &$+0.38$&$+0.01$&+0.47& $+0.34$&$-0.17$&+0.20&+0.47 &9\\
 BD+39 4926&$-2.85 $&$+2.45$ &$+2.75$&$+1.35 $&$+1.15$&$+2.95$ &$-0.75$ &$+1.60$ &$+0.25$&10\\
            \noalign{\smallskip}
            \hline
            \noalign{\smallskip}
\end{tabular}

References:  1 -- This work, 2 -- Van Winckel (1995) ,
 3 --Klochkova (1995),
 4 -- Van  Winckel et al. (1996), 5 -- Lambert et al. (1988), 6 -- Luck et al. (1990),  7 -- Luck \& Bond (1984), 8 -- Bond \& Luck (1987), 9 -- Reddy (1996), 10 -- Kodaira (1973), 11 --  Van Winckel \& Reyniers (2000).

\end{center}
\end{table*}

In terms of evolution, we could distinguish three groups among our sample 
stars. First, HD 158616, HD 172324, HD 172481, and HDE 341617 
show clear indications of being post-AGB stars, these are shown in panel {\it c}
of Fig. 5. The post-AGB  model sequences of Bl\"ocker (1995) suggest that all
four stars had initial ZAMS masses larger than 7 $M_{\odot}$ and remnant or core
mass of $\sim 1 M_{\odot}$. HD 158616 and HD 172481 
might be starting their trajectory towards the
white dwarfs region, i.e. they are near the zero age of central star evolution 
(Bl\"ocker 1995), while HD 172324 is a bit more evolved. Since the evolution
in this region of the H$-$R diagram is very fast, they are all expected to
become planetary nebulae within a few hundreds of years. HDE 341617
has been found to be in the early stages of PN (Arkhipova et al. 1999;
Parthasarathy et al. 2000). Our remnant mass estimate of $\sim 1 M_{\odot}$
is substantially larger than 0.7 $M_{\odot}$ estimated by Arkhipova et al. (1999)
from the rate of temperature evolutionary change and a comparison with the theoretical rates from Bl\"ocker's (1995) models.
Our value is a consequence of the spectroscopic determination of $T_{\rm eff}$ = 23000 K and hence log $(L/L_{\odot})$ = 4.6 (Schmidt-Kaler 1982) as well as a comparison with the luminosities of Bl\"ocker's (1995) models. While the 
 Arkhipova et al.'s (1999) estimation is quite convincing the luminosity of the
adopted model $(L/L_{\odot})$ = 4.0 seems too low as it  would imply 
$T_{\rm eff}$ $\sim$ 18000 K for bright giant star of luminosity class II. Such low
a temperature is not supported by the detailed spectroscopic analyses. Thus
an independent estimate of the luminosity seems necessary to settle the core
mass of this star.

In the second evolutionary group we include the stars HD 725, HD 218753
and HD 331319. These are all moderately iron-deficient but otherwise
show nearly solar abundances. The  heliocentric
  radial velocities for HD 218753 and HD 331319 are small. 
They are most likely young massive disk supergiants or bright giants 
that have gone through some nuclear processing. This is suggested by the C depletion and Na enhancement, indicating the effect of CN processing and post first dredge-up stage.
HD 725 is an interesting star since three Y II and one Ba II lines indicate
mild enhancement of $s$-process elements. Since we count on very scarce number of lines of $s$-process elements, 
we can only say that the star shows signs of evolution beyond RGB. 

\begin{figure*}
\centering
\mbox{\epsfxsize=6.0in\epsfbox{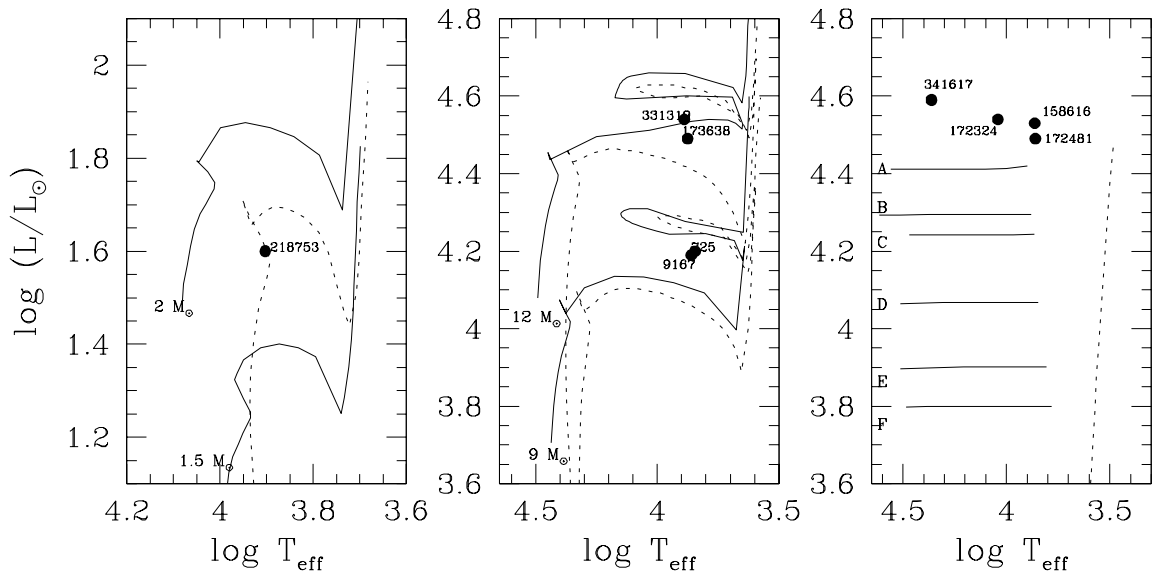}}
\caption{H-R diagram showing the positions of sample stars
along evolutionary tracks (continuous lines) and isochrones (dashed lines).
$T_{\rm eff}$ and log $(L/L_{\odot})$ are those in Table 2.
 The evolutionary tracks are
from Schaller et al. (1992) and all isochrones from Bertelli et al. (1994). (a) The star
HD 218753 has $M \sim 1.5-2.0 M_{\odot}$ and age of $7.9 \times 10^8$ yr. (b)
The two isochrones shown correspond to $3.16\times 10^7$ and $2.0\times 10^7$ yr. See text for
discussion. (c) The post-AGB models are from Bl\"ocker 1995 for ZAMS
 mass and core mass combinations ($M_{zams},M_{\rm H}$); A: = (7$M_{\odot}$,0.940$M_{\odot}$), B:
= (3$M_{\odot}$,0.836$M_{\odot}$), C: = (5$M_{\odot}$,0.836$M_{\odot}$), D: = (4$M_{\odot}$,0.696$M_{\odot}$), E: = (3$M_{\odot}$,0.625$M_{\odot}$), F: = (3$M_{\odot}$,0.605$M_{\odot}$). The isochrone shown corresponds to $2.5\times 10^8$ yr. See text for discussion.}
\end{figure*}

The tracks and isochrones on Figs. 5a and 5b suggest $M \sim 1.5-2.0 M_{\odot}$ and age of $7.9 \times 10^8$ yr for HD 218753 and 
$M \sim 11 M_{\odot}$ and age of $2.2 \times 10^7$ yr for HD 331319. 
 From tracks on Fig. 5b we estimate  a mass of 
$M \leq 9 M_{\odot}$ and age of $2.5 \times 10^7$ yr. Its rather 
large heliocentric radial velocity of $-$57 km~s$^{-1}$ calls
  for  attention, however, a calculation of the
galactocentric motion indicates that the star moves on the galactic plane and 
has a mildly
eccentrical galactic orbit. 
Its radial velocity could owe  its origin to
pulsations and/or orbital motions, although variability has not been reported.

The last two stars in our sample, HD 9167 and HD 173638, 
 display, with few exceptions, solar abundances and no signs of 
nuclear processing . They are probably evolving very near the giant branch. 
The estimated masses and ages from Fig. 5b are respectively 
$M \sim 10 M_{\odot}$ and $2.6 \times 10^7$ yr  and 
$M \sim 12 M_{\odot}$ and $1.7 \times 10^7$ yr. The 
only peculiarity of HD 9167 is its high 
radial velocity of $-$45.7 km~s$^{-1}$ however, like HD 725, 
its galactocentric orbit seems to be on the galactic plane and mildly 
eccentric. Also the possibility, that their observed radial 
velocities are attained from pulsation and/or 
orbital motion cannot be discarded.

 It should be noted that although HD 172481 and HD 158616  are post-AGB 
 stars, they do not show the effect of selective removal
 of condensable elements such as Fe and Sc, observed in some well-known post-AGB stars like 
 HR 4049, and HD 52961 and RV Tau stars of subclass B (Giridhar, Lambert \& Gonzalez 2000 and references therein).   
While studying a sample of RV Tau stars, these authors had
 noticed a strong dependence on temperature for the selective removal
 of refractory elements to occur. The effect is very prominent at temperature range 5500 to 6000 K 
and declines for lower temperatures. For stars cooler than 5000 K 
the effect was barely 
perceptible. At temperatures higher than 7000 K, we expect the effect to be larger. It is indeed true for HR 4049 (Lambert et al. 1988) which has 
$T_{\rm eff}$ = 7500 K, i.e. similar to HD 158616 and HD 172481. 
However, for these
two stars we did not see any indication of dust condensation and subsequent
removal of grain-forming elements.
HD 158616 is a carbon-rich post-AGB star similar to HD 56126 
(Klochkova 1995) and 
HD 187785 (Van Winckel et al. 1996) also showing significant
enhancement of $s$-process elements.

Stars like HR 4049, HD 52961 and RV Tau stars of subclass B show C/O $\leq$ 1 and mild $s$-process enhancement. 
As a matter of fact,
most stars showing abundance peculiarities caused by dust condensation
have C/O $\leq$ 1.
Stars HR 4049, HD 44179, HD 46703, HD 52961 and BD +39$^{o}$ 4926  possibly
belong to this subgroup. For these objects, since Fe gets locked in grains,  
[S/H] is considered 
a better indicator of metallicity. For these stars  [S/H]
ranges between +0.1 to $-$1.0 dex with a mean around $-$0.4 dex. In other words,
they are mildly metal-deficient. Carbon-rich post-AGB stars with enhanced $s$-process elements, like HD 158616, HD 56126 and HD 187785 have [Fe/H] (which
would be a true reflection of their metallicity since these stars are not affected by dust condensation) in the range $-$0.4 to $-$1.0 dex. These values
are not radically different from those found for the subgroup having 
 dust-grain condensation and 
C/O $\leq$ 1. We, therefore, do not visualize large differences in 
their ages though the O-rich phase in the AGB is expected to precede
the C-rich phase.

The abundances of hot post-AGB
 stars studied by Conlon et al. (1993a) and McCausland et al. (1992)
 bear close resemblance to HD 172324. The hot post-AGB stars show strong deficiency of carbon and significant oxygen enrichment. These stars probably belong
 to a subgroup of post-AGB star that have evolved without experiencing
 third dredge-up.  This carbon deficiency is also found in the proto-Planetary
Nebula HDE 341617 (see Table 13).
Caution is however needed with C II spectra since they are known to show large
non-LTE effects (Eber \& Butler 1988; Takeda \& Takada-Hidai 1994). 
McCausland et al. (1992) have discussed at length two scenarios to explain
 the carbon deficiency. HBB occuring during interpulse phase could cause
 the production of $^{14}$N at the expense of $^{12}$C. However, overabundance 
 of He like the one found in the SMC planetary nebula SMP 28 is not
 evident for HD 172324 and HD 341617 to make HBB the sole mechanism responsible
 for carbon deficiency. Another possibility suggested by 
 McCausland et al. (1992) that the carbon deficiency might be inherent
 to the precursor itself  is quite attractive. To substantiate their 
 argument they pointed out the carbon-poor stars           
HR 4912 and HR 7671 as possible precursors to more evolved carbon-poor 
hot post-AGB stars. HR 4912 was included in our recent work
 and we found [C/H] = $-$1.27 (Giridhar et al. 1997) in good agreement 
 with [C/H] of $-$1.15 found by Lambert, Luck \& Bond (1983). HR 7671 has 
 [C/H] of $-$1.53 (Luck et al. 1990). It seems therefore that HD 172324
and  HDE 341617  might form a special carbon-poor post-AGB stars evolutionary
sequence.  Search for carbon-poor objects in all
 temperature ranges may help in finding the precursors or successors of these
 objects.
                                           
\section {Conclusions}

 We have found a new post-AGB star HD 172481 for which abundance
 analysis had not been carried out before, however the referee pointed out
in a rather late stage of revision,
the then unpublished work by Reyniers \& Van Winckel (2001) 
 where an independent 
abundance analysis has been carried out. We have highlighted a comparison
of results in section 5.5 and found both analyses to be in fairly good agreement. 
We have done a more 
 complete analysis of HD 158616. This star can be now considered a 
 post-AGB star beyond any doubt. Among the post-AGB stars, the number of
 stars showing C/O $\sim 1$ or greater  and also enhancement of $s$-process
 elements are very few. There are clear indications of stars having
 experienced the third dredge-up. The stars HD 158616 and HD 172481
 belong to this important class. 
 C/O of  HD 172481 might have been prevented from exceeding one
 by HBB. The same may be responsible for large Li
 abundance but other possibilities like binarity cannot be ruled out.
 A long-term light and radial velocity monitoring of this object
 is planned for the future.

 We found a very likely  hot post-AGB candidate
 in HD 172324 but will feel more confident of its status after
 important elements like N are included and a more comprehensive study is made.
 Continuous monitoring of its spectrum in the H$_\alpha$ region is required
 to detect activity possibly related to stellar pulsations. 

 HD 218753 and HD 331319 have passed the giant branch and are in the 
He-core and H-shell burning stages. HD 9167 and HD 173638 essentially show solar
 abundance. HD 725 and HD 9167 show large radial 
velocities, however their galactocentric orbits are on the galactic plane and
mildly eccentric, thus they are most likely  massive and young disk stars.
However, their large radial velocity could also be due to  pulsations and/or orbital motions. 

In the proto-Planetary Nebula HDE 341627 the He lines show two velocity 
components  possibly indicating velocity stratification. The emission lines 
 appear to have weakened since 1993.
     
\acknowledgements
AAF acknowledges Commission 38 of the IAU for
a travel grant and the Indian Institute of Astrophysics for hospitality and 
financial support.
 We thank David Yong for getting us  two spectra  of HD 172481.
 We are also indebted to Ms. T. Sivarani for her help in identifying 
 the lines for HDE 341617. We are grateful  to the referee, Dr. R. Gallino, for helpful discussions and suggestions.
  This project has been supported
at different stages by grants from DGAPA-UNAM (IN113599)
and CONACyT (Mexico) (E130.2060) and CNRS (France). 
%

\end{document}